\documentclass{emulateapj}
\usepackage{gensymb}
\usepackage{upgreek}
\usepackage{color}

\begin{document}

\shortauthors{Esplin \& Luhman}
\shorttitle{Survey for Brown Dwarfs in Taurus and Perseus}

\title{A Survey For Planetary-mass Brown Dwarfs in the Taurus and Perseus
Star-forming Regions\altaffilmark{1}}

\author{
T. L. Esplin\altaffilmark{2} and 
K. L. Luhman\altaffilmark{2,3} 
}

\altaffiltext{1}{Based on observations made with the NASA Infrared Telescope
Facility, Gemini Observatory, Pan-STARRS1, 2MASS, UKIDSS, SDSS, {\it Gaia},
{\it WISE}, and the {\it Spitzer Space Telescope}, which is operated by the
Jet Propulsion Laboratory, California Institute of Technology under a contract
with NASA.}
\altaffiltext{2}{Department of Astronomy and Astrophysics, The Pennsylvania
State University, University Park, PA 16802; taran.esplin@psu.edu.}
\altaffiltext{3}{Center for Exoplanets and Habitable Worlds,
The Pennsylvania State University, University Park, PA 16802.}

\begin{abstract}

We present the initial results from a survey for planetary-mass brown dwarfs
in the Taurus star-forming region. We have identified brown dwarf candidates
in Taurus using proper motions and photometry from several ground- and
space-based facilities. Through spectroscopy of some of the more promising
candidates, we have found 18 new members of Taurus.
They have spectral types ranging from mid M to early L and they include the
four faintest known members in extinction-corrected $K_s$, which should
have masses as low as $\sim4$--5~$M_{\rm Jup}$ according to evolutionary models.
Two of the coolest new members (M9.25, M9.5) have mid-IR excesses that indicate
the presence of disks. Two fainter objects with types of M9--L2 and M9--L3
also have red mid-IR colors relative to photospheres at $\leq$L0, but
since the photospheric colors are poorly defined at $>$L0, it is unclear
whether they have excesses from disks. We also have obtained spectra of 
candidate members of the IC~348 and NGC~1333 clusters in Perseus that
were identified by \citet{luh16}. Eight candidates are found to be probable
members, three of which are among the faintest and least-massive
known members of the clusters ($\sim5$~$M_{\rm Jup}$).

\end{abstract}

 
\section{Introduction}

The identification of large samples of planetary-mass brown dwarfs
($<15$~$M_{\rm Jup}$) is important for measuring the minimum mass of the
initial mass function and for helping to interpret observations of directly
imaged planetary companions.
This is most easily done in the nearest young clusters and associations 
($\lesssim10$~Myr, 100--300~pc) given the sensitivities of existing telescopes
and the predicted luminosities of brown dwarfs as a function of age
\citep{bur97,cha00}.
The Taurus and Perseus star-forming regions are two of the most appealing
targets for a brown dwarf survey because of their
proximity \citep[$d=140$ and 300~pc;][references therein]{tor12,sch14} and 
the relatively large sizes of their stellar populations
\citep[N$\sim$400 and 800,][]{ken08,bal08}.
\citet{luh16,luh17} have provided recent summaries of previous surveys
for members of Taurus and the two richest clusters in Perseus, IC~348
and NGC~1333 (see also \citet{kra17} for Taurus).
The census of each region contains several dozen objects with spectral
types indicative of brown dwarfs ($>$M6), most of which were identified
with optical and infrared (IR) photometry. 

We have begun a new search for brown dwarfs in Taurus that improves upon
previous surveys of the region in terms of both coverage and depth.
This work is based on photometry and astrometry from wide-field images
collected by the Two Micron All Sky Survey \citep[2MASS,][]{skr06},
the {\it Spitzer Space Telescope} \citep{wer04}, the United Kingdom Infrared
Telescope Infrared Deep Sky Survey \citep[UKIDSS,][]{law07}, Pan-STARRS1
\citep[PS1;][]{kai02,kai10}, the Sloan Digital Sky Survey 
\cite[SDSS,][]{yor00,fin04}, {\it Gaia} \citep{per01},
and the {\it Wide-field Infrared Survey Explorer} \citep[{\it WISE},][]{wri10}.
In addition, we have continued the survey of IC~348 and NGC~1333 from 
\citet{luh16} by obtaining spectra of candidates for planetary-mass brown
dwarfs from that study.
In this paper, we apply updates to the census of previously known members
of Taurus from \citet{luh17} (Section~\ref{sec:mem}), identify candidate
members of Taurus using color-magnitude diagrams (CMDs) and proper motions
(Section~\ref{sec:can}), present spectroscopic classifications of an initial
sample of promising candidates (Section \ref{sec:spec}), and check the new
members for mid-IR excess emission that would indicate the presence of
circumstellar disks (Section \ref{sec:disks}).
We then present spectroscopy for several candidate members of
IC~348 and NGC~1333 from \citet{luh16} (Section~\ref{sec:pers}).

\section{Previously Known Members of Taurus}
\label{sec:mem}

Before presenting our survey for brown dwarfs in Taurus, we describe our
adopted list of previously known members.
A recent compilation of members was presented by \citet{luh17}.
Since that study, \citet{kra17} and \citet{bes17} have proposed additional
members of Taurus. In this section, we examine those candidates to determine
whether to add them to the list from \citet{luh17}. We also reject a few
stars from the latter that do not appear to be members.

\subsection{Candidate Members from \citet{kra17}}
\label{sec:kra17}

\citet{kra17} compiled diagnostics of membership in the Taurus star-forming
region for 396 candidate members that had been identified in previous surveys.
They selected candidates that lacked evidence of disks in earlier
studies\footnote{One of these stars, 2MASS J05080816+2427150, was classified
as diskless by \citet{esp14}, but we find that it does show evidence of a disk
(Section~\ref{sec:disks}).} and that were within a
large area extending well beyond the Taurus clouds that was defined by
right ascensions ($\alpha$) of $3^{\rm h}50^{\rm m}$--$5^{\rm h}40^{\rm m}$
and declinations ($\delta$) of $14\arcdeg$--$34\arcdeg$ (J2000). 
The diagnostics consisted of positions in the Hertzsprung-Russel diagram,
lithium absorption, gravity-sensitive spectral features, H$\alpha$ emission,
proper motions, and radial velocities.
A given candidate was treated as a member of Taurus if it appeared to be a
pre-main-sequence star based on the first four measurements (when available)
and if its kinematics were similar to those of the previously known members.
Some candidates lacked sufficient data for definitive assessments of their
membership.
\citet{kra17} concluded that 218 of the 396 candidates were confirmed or likely
members, 82 of which were absent from the compilation of members in \citet{luh17}.
Most of those 82 stars are older (10--30~Myr) and more widely distributed
than the previously known members ($\sim1$~Myr), so \citet{kra17} proposed
that they represent an earlier generation of star formation that is related
to the Taurus cloud complex. Although it was not included in their sample,
\citet{kra17} noted St~34 as an additional example of an intermediate-age
pre-main-sequence star that appears to be co-moving with the known Taurus
members, satisfying their criteria for membership.

For four stars adopted as members by \citet{luh17}, \citet{kra17} found
insufficient data for membership assessment (V410~Anon~25, V410~X-ray~4)
or proper motions that appeared inconsistent with membership
(V410~X-ray~5a, LH~0419+15).
We treat the first two stars as members because they are too highly reddened
for foreground stars ($A_V\sim25$ and 19), are too bright for background
dwarfs, have 2.3~\micron\ CO bandheads that are too weak for background giants
\citep{luh17}, and have proper motions that are consistent with membership
(Section~\ref{sec:pm:irac}).
Both of the other two stars, V410~X-ray~5a and LH~0419+15, are 
very young ($\lesssim10$~Myr) based on the gravity-sensitive features in their
spectra \citep[][references therein]{luh17}. Our measured proper motions
for V410~X-ray~5a also support membership (Sections~\ref{sec:pm:irac},
\ref{sec:pm:pan}). However, LH~0419+15 is unlikely to be a member
based on the proper motion data from
\citet{har99} and Section~\ref{sec:pm:pan} ($\mu_{\alpha},
\mu_{\delta}=0.6\pm4.1$, $-5.5\pm5.4$~mas~yr$^{-1}$).
It is also fainter than any other known member near its spectral
type (M6--M7). Therefore, we no longer adopt LH~0419+15 as a member.

\begin{figure}[h]
\centering
\includegraphics[trim = 0mm 0mm 0mm 0mm, clip=true, scale=0.8]{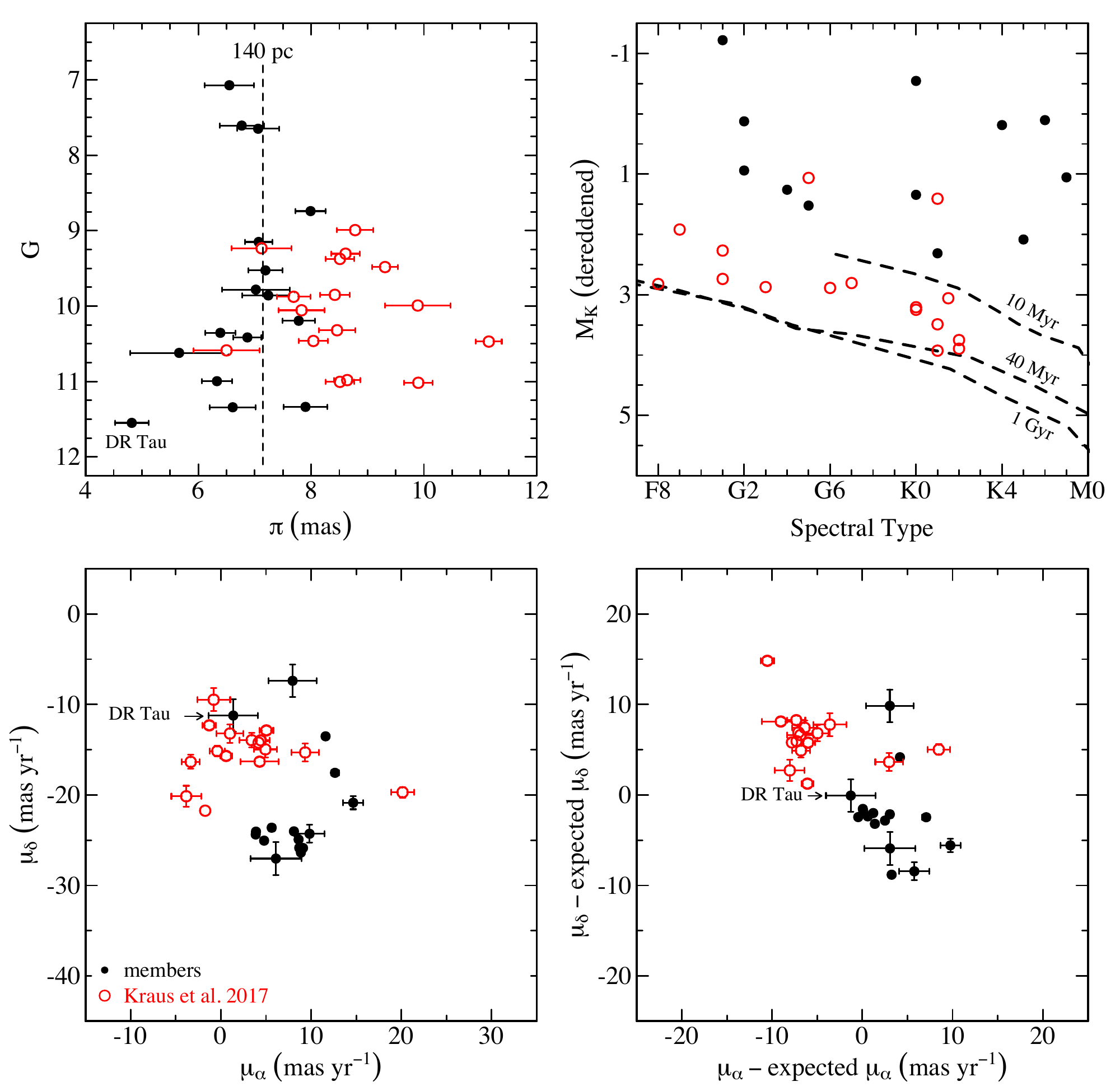}
\caption{
Previously known members of Taurus adopted by \citet{luh17} (filled circles)
and additional candidate members from \citet{kra17} (open circles) that
have {\it Gaia} parallaxes and proper motions and  are within the
boundaries of our survey ($\alpha=4^{\rm h}$--$5^{\rm h}10^{\rm m}$,
$\delta=15$--$31\arcdeg$). Top left: {\it Gaia} magnitude versus parallax. 
Most of the candidates have parallactic distances of 100--120~pc,
placing them in the foreground of the known members.
One of the known members, DR~Tau, exhibits a discrepant parallax
relative to the other members.
Top right: extinction-corrected $M_K$ versus spectral type with
model isochrones from \citet{bar15}, which indicate ages of 10--40~Myr
for most of the candidates.
Bottom left: {\it Gaia} proper motions.
Bottom right: Offsets of the proper motions relative to the values expected
for the positions and parallaxes of the stars assuming the mean space velocity
of Taurus members \citep{luh09b}. The spread in motions due to projection
effects should be reduced in these offsets. In both of the bottom diagrams,
most of the candidates have motions that are distinct from those of
the known Taurus members.
}
\label{fig:pi}
\end{figure}

The field that we have considered for our survey of Taurus
($\alpha=4^{\rm h}$--$5^{\rm h}10^{\rm m}$, $\delta=15$--$31\arcdeg$,
Section~\ref{sec:can}) is smaller than that from \citet{kra17}, but it
still encompasses all of the Taurus clouds and the previously adopted members
from \citet{luh17}. Our field contains 56 of the 82 proposed members
from \citet{kra17} that were absent from \citet{luh17}. One of those 56 stars,
XEST 08-014, was spectroscopically classified as a field dwarf by \citet{luh09b}.
A second star, HBC~407, was rejected by \citet{luh17} because its proper
motion was inconsistent with membership.
Among the remaining 54 candidates from \citet{kra17}
within our survey field, 16 have measurements of parallaxes and proper motions
in the first data release of {\it Gaia}.
To assess the membership of those 16 stars, we plot them
in diagrams of {\it Gaia}
magnitude ($G$) versus parallax, $\mu_{\delta}$ versus $\mu_{\alpha}$, and
extinction-corrected $M_K$ versus spectral type in Figure~\ref{fig:pi}.
The latter diagram is based on $K_s$ from the 2MASS Point Source Catalog,
the extinctions and spectral types adopted by \citet{kra17}, and the
{\it Gaia} parallactic distances. For comparison, we also show
the 16 known members from \citet{luh17} that have {\it Gaia}
parallaxes and proper motions.
Their spectral types and extinctions are taken from \citet{luh17}.
For a given space velocity, a star's proper motion varies with position
on the sky and distance. To reduce these projection effects, we have
subtracted from the proper motion of each star the motion expected
for the mean space velocity of known Taurus members \citep{luh09b}
and the $\alpha$, $\delta$, and parallax of the star. 
A diagram of the resulting offsets is included in Figure~\ref{fig:pi}.

Most of the known members of Taurus in Figure~\ref{fig:pi} have parallactic
distances near the value of $\sim$140~pc that has been previously measured
for a small subset of members \citep{wic98,loi05,tor07,tor09,tor12}.
The primary exception is DR~Tau, which would appear to be located 
behind Taurus according to {\it Gaia} ($207^{+14}_{-12}$~pc).
In comparison, most of the candidates have closer distances of 100--120~pc,
which was noticed by \citet{kra17} during their examination of the same
{\it Gaia} parallaxes. 
The diagram of $M_K$ versus spectral type in Figure~\ref{fig:pi}
indicates that most of the candidates are older than the known members,
appearing between the isochrones for 10 and 40~Myr from \citet{bar15}.
In the proper motion diagrams, most of the candidates are distinct from
the known members; the two populations differ by $\sim10$~mas~yr$^{-1}$.
2MASS~J04244815+2643161 is the only candidate for which both its
proper motion and parallax are similar to those of the known members.
In addition, it is one of the two candidates that appear above the 10~Myr
isochrone in Figure~\ref{fig:pi} and it is located near known members.
As a result, we choose to adopt it as a member of Taurus.
We note that the proper motion measurement for St~34 from \citet{zac17}
($\mu_{\alpha}, \mu_{\delta}=5.8\pm1.4$, $-12.7\pm1.3$~mas~yr$^{-1}$)
coincides with the group of candidates from \citet{kra17} rather than
the known members (Figure~\ref{fig:pi}). \citet{har05st} proposed that St~34
is $\sim$30--40~pc closer than Taurus, which would also place it in the same
range of distances as those candidates.

\begin{figure}[h]
\centering
\includegraphics[trim = 0mm 0mm 0mm 0mm, clip=true, scale=0.7]{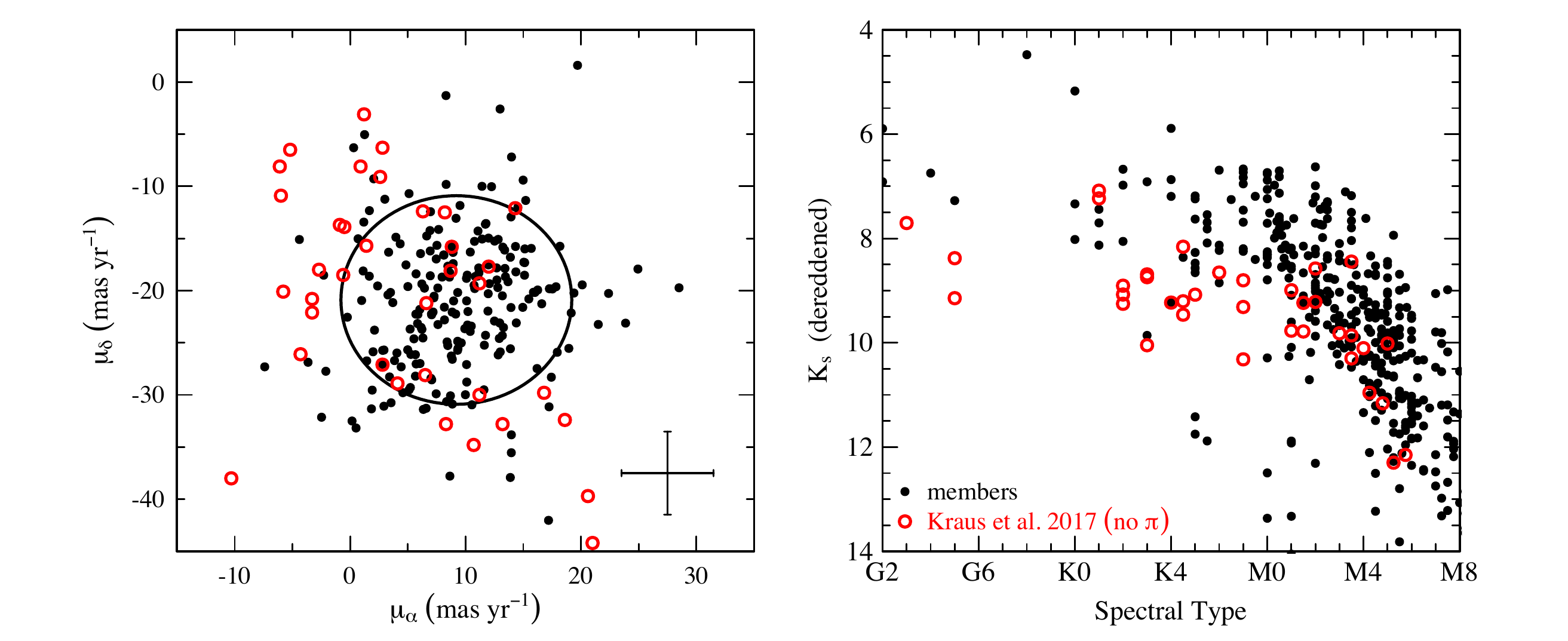}
\caption{
Previously known members of Taurus adopted by \citet{luh17} (filled circles) and
additional candidate members from \citet{kra17} (open circles) that lack
{\it Gaia} parallaxes (i.e., not in Figure~\ref{fig:pi}) and are within the
boundaries of our survey ($\alpha=4^{\rm h}$--$5^{\rm h}10^{\rm m}$,
$\delta=15$--$31\arcdeg$). Left: proper motions measured with astrometry 
from 2MASS/{\it Gaia} or 2MASS/PS1.
For reference, we have marked the threshold that we have applied to proper
motions from these catalogs for our survey (large circle).
The typical errors for these data are indicated.
Right: extinction-corrected $K_s$ versus spectral type.
Most of the members that are unusually faint
for their spectral types are occulted by circumstellar material and seen in 
scattered light, or they exhibit mid-IR excess emission indicating the 
presence of a disk, making scattered light from an occulting disk a
possibility.
None of the candidates show mid-IR excesses, so those that appear below the 
Taurus sequence are likely older or more distant than the known members.
}
\label{fig:nopi}
\end{figure}

Parallaxes are not available for 38 of the 54 candidates from \citet{kra17}
in our survey field. For 36 of those 38 stars, proper motions can be
measured with astrometry from 2MASS and {\it Gaia} or 2MASS and PS1.
In Figure~\ref{fig:nopi}, we plot the 2MASS/{\it Gaia} motions for those
36 candidates when available, and otherwise show the 2MASS/PS1 measurements.
The 38 candidates also are plotted in a diagram of extinction-corrected $K_s$
versus spectral type in Figure~\ref{fig:nopi}. In both diagrams, we have included
the previously known members from \citet{luh17}. 
We have marked the 10~mas~yr$^{-1}$ radius threshold in the proper motion
diagram that we apply to 2MASS/{\it Gaia} motions in our survey for new members 
(Section~\ref{sec:2m.gaia}).
That criterion is satisfied (at 1~$\sigma$) by 24 of the 36 candidates from
\citet{kra17} whose proper motions are shown in Figure~\ref{fig:nopi}. 
Because the errors in these proper motions are larger than those from {\it Gaia},
we are unable to determine whether the candidates in Figure~\ref{fig:nopi} form
a well-defined, distinct population like that found for the candidates with
{\it Gaia} motions in Figure~\ref{fig:pi}.

In the diagram of $K_s$ versus spectral type in Figure~\ref{fig:nopi}, many
of the candidates are fainter than the sequence of the known members, indicating
that they have older ages if they are near the distance of the Taurus clouds. 
Only five candidates satisfy our proper motion threshold,
appear within the sequence of known members in $K_s$ versus spectral type,
and exhibit evidence of youth among the diagnostics from \citet{kra17}.
Two of them, 2MASS J05080816+2427150 and 2MASS J04355683+2352049, are
near known members, and the former has red mid-IR colors that indicate
the presence of a disk (Section~\ref{sec:disks}), so we add them to our
census of Taurus.
For the other three stars that are far from members, 2MASS J04091700+1716081,
2MASS J04515424+1758280, and 2MASS J04525015+1622092, we defer an assessment
of their membership until the next data release for {\it Gaia}, which should
provide measurements of their proper motions and parallaxes.

In the preceding discussion, among the candidates from \citet{kra17} that
were absent from the compilation of known members in \citet{luh17}, we
adopted three stars as Taurus members because they appear to have similar
ages, kinematics, and distances (when available) as those known members.
Most of the other candidates are older than the known members, and many have
kinematics or distances that are noticeably different.
In particular, the subset of candidates with {\it Gaia} parallaxes and
proper motions are distinct from the known members in the {\it Gaia} data.
As a result, we contend that those candidates and the known members 
should not be treated as constituents of the same stellar population, even if
their origins are related. 

Any relationship between the older candidates from \citet{kra17}
and the known Taurus members may be tenuous.
\citet{kra17} suggested that their candidates formed from either the
current Taurus clouds or previously existing clouds that have dispersed.
The former scenario is unlikely given that 1) the lifetimes of molecular
clouds \citep[$\sim$1--2 Myr,][]{har01,har12} are much shorter than the ages
of those stars (10--40~Myr), 2) the candidates are not isotropic relative
to the known members in the {\it Gaia} proper motions, and 3) the candidates
are systematically offset relative to the known members in the {\it Gaia}
parallaxes. Instead, it is much more likely that the candidates originated
in clouds that are no longer present. 
\citet{kra17} proposed that those earlier clouds were closely associated
with the current Taurus clouds such that they together represented a long-lived
star-forming complex. However, the differences in {\it Gaia} proper motions and
parallaxes for candidates and known members (when available) indicate that
the clouds that produced the former may have been quite far from the gas
that would eventually become the Taurus clouds.
For instance, the offset of $\sim$10~mas~yr$^{-1}$ in the {\it Gaia} proper
motions of the two populations corresponds to a relative drift of nearly
$30\arcdeg$ over 10~Myr.
It is possible that the natal clouds of the candidates from \citet{kra17}
and the known Taurus members have no relationship beyond the fact that,
like a number of other molecular clouds, they both formed just beyond the edge
of the Local Bubble.

\subsection{Candidate Members from \citet{bes17}}
\label{sec:best}

\citet{bes15,bes17} recently used photometry from {\it WISE}
and PS1 to search for brown dwarfs in the solar neighborhood
that are near the L/T transition.
They avoided areas of high extinction like the dark clouds in Taurus, but
their survey did include the outskirts of Taurus, where they uncovered
two late-type dwarfs, PSO J060.3+25 (2MASS J04011678+2557527) and
PSO J077.1+24 (2MASS J05082480+2422518). Both objects were classified as
young L dwarfs and proposed as new members of Taurus.

\citet{bes17} used near-IR spectroscopy to measure spectral types of L1 and
L2 for PSO J060.3+25 and PSO J077.1+24, respectively.
To obtain types that have been measured in the same system as the known
late-type members of Taurus, we have classified the spectra from \citet{bes17}
with the M/L standard spectra from \citet{luh17}.
The latter were constructed from spectra of M-type members of Taurus and
other young populations ($\lesssim$10~Myr) that have optical spectral types
measured with the methods from \citet{luh99} and spectra of the
youngest field L dwarfs that have optical types derived
with the scheme from \citet{cru09} and \citet{kir10}.
When comparing each PS1 object to a standard at a given spectral type, the
latter was artificially reddened to match the spectral slope of the former
using the extinction law from \citet{car89} (only positive values of
extinction were allowed). 

\begin{figure}[h]
\centering
\includegraphics[trim = 0mm 10mm 0mm 0mm, clip=true, scale=0.9]{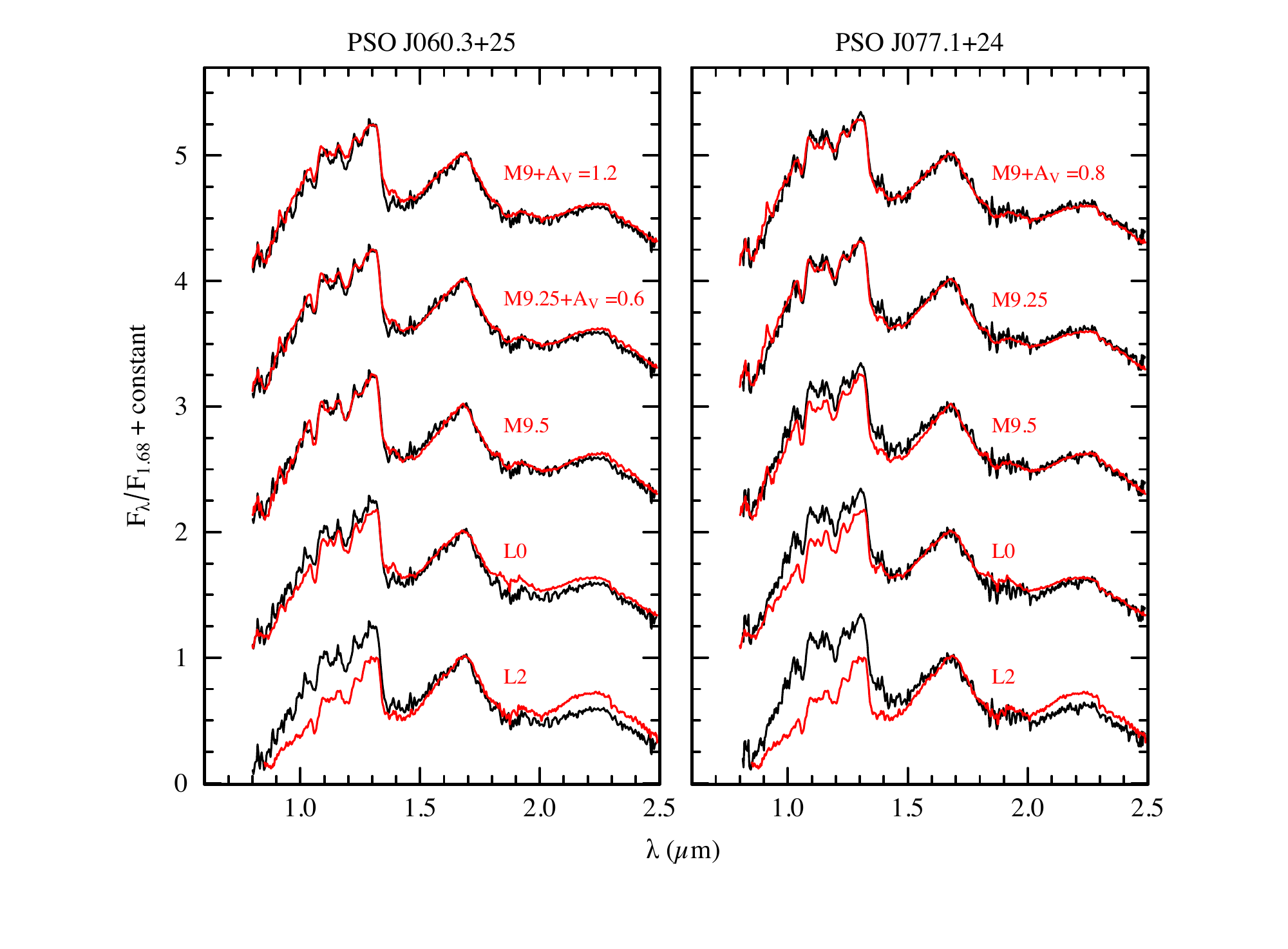}
\caption{
Near-IR spectra of PSO~J060.3+25 and PSO~J077.1+24 from \cite{bes17} (black
lines), which were proposed as members of Taurus and were classified as L1 and
L2 in that study, respectively. Those data are compared to standard spectra
from M9--L2 for ages of $\lesssim10$~Myr \citep[red lines,][]{luh17}.
If PSO~J060.3+25 or PSO~J077.1+24 was redder than a given standard spectrum,
the latter was artificially reddened to match the slope of the former.
If they are members of Taurus, both objects would have types of M9.25 
according to these standards. 
However, we find that neither object shows strong evidence of membership
in Taurus (Section~\ref{sec:best}).
}
\label{fig:best}
\end{figure}

In Figure~\ref{fig:best}, we compare the data for
PSO~J060.3+25 and PSO~J077.1+24 to standard spectra between M9 and L2 with
the best-fitting reddenings.
We find that the spectra of PSO~J060.3+25 and PSO~J077.1+24 are best
matched by M9.25 with $A_V=0.6$ and M9.25 with no extinction, respectively.
Both objects are much bluer than our standard spectra for young L dwarfs
as well as the spectrum from \citet{bow14} for Taurus member
2MASS~J04373705+2331080, whose
optical spectrum exhibits a type of L0 and little extinction \citep{luh09b}.
As discussed by \citet{bes17}, PSO~J060.3+25 and PSO~J077.1+24 are clearly
young based on the gravity-sensitive features in their spectra.
However, PSO~J060.3+25 has an $H$-band continuum that is slightly less
triangular and an FeH feature (0.99~\micron) that is somewhat stronger
than our young standards, indicating that it may be older than the objects
associated with the Taurus dark clouds ($>10$~Myr).
PSO~J077.1+24 also has stronger Na~I absorption at 2.2~\micron\ than the
standards, which would suggest an older age as well, although the S/N is
low near that feature.
We note that our classifications are valid only if the PS1 objects
have ages similar to those of known Taurus members. If they are older, then
one would need to perform the classifications with older standards, which
could produce slightly later spectral types than those we have derived
with $<10$~Myr standards given that near-IR spectra are bluer and the
steam bands are more shallow at older ages for a given optical M/L type.

\begin{figure}[h]
\centering
\includegraphics[trim = 0mm 0mm 0mm 0mm, clip=true, scale=0.8]{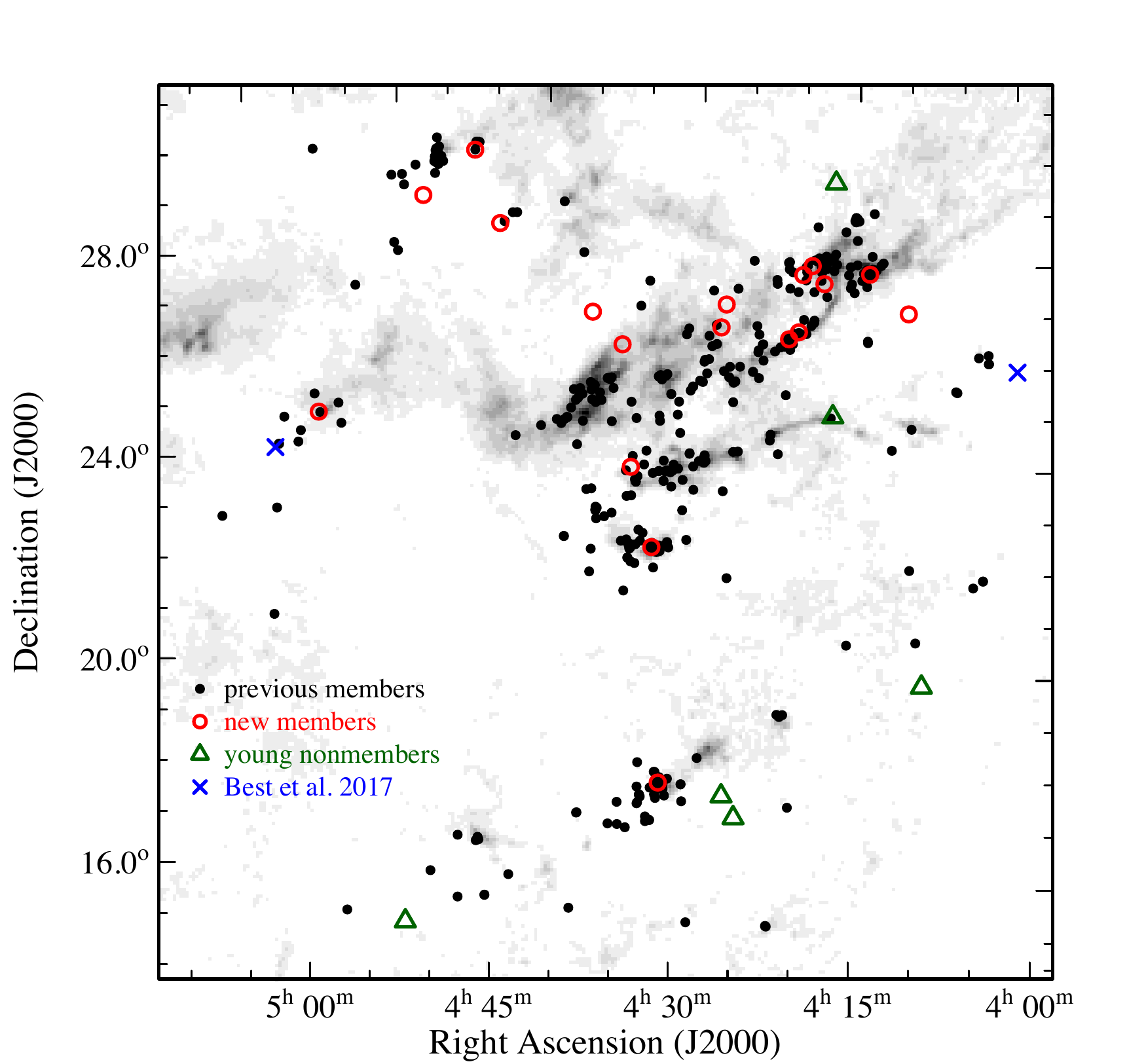}
\caption{
Spatial distribution of previously known members of Taurus (filled circles),
new members from this work (open circles),
young objects found in our survey that do not appear to be members 
(open triangles), and two candidate members from \cite{bes17} (crosses).
The dark clouds in Taurus are displayed with a map of extinction
\citep[gray scale;][]{dob05}.
}
\label{fig:new}
\end{figure}

PSO~J060.3+25 and PSO~J077.1+24 are located on the western and eastern edges
of Taurus, respectively, as shown in Figure~\ref{fig:new}.
PSO~J060.3+25 is also on the western periphery of the Pleiades open
cluster \citep[125~Myr,][]{sta98}, and in fact was identified
as a candidate member of the Pleiades by \citet{sar14} and \citet{bou15}
based on its photometry and proper motion.
\citet{bes17} found that the proper motion of PSO~J060.3+25 from
\citet{bou15} was consistent with membership in either Taurus or the Pleiades,
and that the PS1 motion for PSO~J077.1+24 was consistent with membership
in Taurus, although the errors in that measurement were fairly large
($\sim12$~mas~yr$^{-1}$).
The positions of PSO~J060.3+25 in diagrams of near-IR magnitudes versus
spectral type from \citet{bes17} coincided more closely with the Taurus
sequence than the Pleiades sequence, so that study favored membership in
Taurus. However, its position agrees better with the Pleiades sequence if our
spectral classification of M9.25 is adopted.
Indeed, both PSO~J060.3+25 and PSO~J077.1+24 are fainter than any known
members of Taurus at M9--M9.5, as illustrated in the diagram of
extinction-corrected $K_s$ versus spectral type in Figure~\ref{fig:kvspec}.
Based on their position in that diagram and their gravity-sensitive spectral
features, we conclude that PSO~J060.3+25 and PSO~J077.1+24 could have
ages of $>$10~Myr and may not be members of the Taurus star-forming population.
Late-type objects with surface gravities intermediate between those of Taurus
members and field dwarfs have been identified in previous surveys of Taurus 
\citep{luhm06,sle06}, some of which were included in the sample of proposed
members from \citet{kra17}. We have found additional late-type objects
with intermediate ages in our survey (Section~\ref{sec:spec}).

\subsection{Updates to the Census from \citet{luh17}}

To construct a list of the previously known members of Taurus, we begin
with the compilation from \citet{luh17}.
The following stars from that list are now rejected because they appear 
unlikely to be members:
L1551-55, RXJ05072+2437, HBC~376, 2MASS J04163048+3037053,
2MASS J04215851+1520145, 2MASS J04374333+3056563, 2MASS J04080782+2807280, 
2MASS J04505356+2139233, ITG~1 (2MASS J04375670+2546229), and LH~0419+15.
The first seven stars are significantly fainter than known Taurus members
near their spectral types. Stars that are occulted by circumstellar material
and seen in scattered light can appear unusually faint, but
six of those seven
stars lack the mid-IR excess emission expected from a circumstellar disk
in photometry from {\it Spitzer} and {\it WISE}. One of those six
stars, 2MASS J04215851+1520145, was identified as a member of Taurus based on a
mid-IR excess in the {\it WISE} All-Sky Source Catalog \citep{esp14}, but newer
data in the AllWISE Source Catalog do not exhibit an excess. 
The  seventh star,
2MASS J04374333+3056563, has an excess only longward of 10~\micron, whereas
an edge-on disk should show excess emission at shorter wavelengths as well.
The eighth
star, 2MASS J04505356+2139233, is rejected because its Li absorption 
is rather weak for a Taurus member near its spectral type \citep{fin10}
and it appears along the lower envelope of the Taurus sequence in $K$ versus 
spectral type.  ITG~1 was classified as a Taurus member based on its 
mid-IR excess emission, which suggested that it harbored a circumstellar disk 
and therefore was likely to be a young star \citep{luh06,fur11}.
However, measurements of its proper motion with {\it Spitzer} and UKIDSS 
($\mu_{\alpha}, \mu_{\delta}=-5.3\pm2.7$, $2.5\pm5.7$ and $-5.1\pm4.7$,
$7.4\pm4.7$~mas~yr$^{-1}$) and from other astrometric catalogs \citep{roe10} 
are inconsistent with membership (Section~\ref{sec:pm}).
ITG~1 is probably in the background of Taurus based on its small proper motion,
possibly a dusty evolved star. The reasons for rejecting LH~0419+15
were described in Section~\ref{sec:kra17}.

As discussed in Section~\ref{sec:kra17}, we have adopted three candidates
from \citet{kra17} as members of Taurus, consisting of 2MASS J04244815+2643161,
2MASS J05080816+2427150, and 2MASS J04355683+2352049.
We have also included 2MASS J04390453+2333199, 2MASS J04354778+2523436, and
2MASS J04225416+2439538 in our list of members. They were
classified as new members by \citet{abe14} based on their proper motions,
spectroscopic indicators of youth, and positions in CMDs.
As mentioned in Section~\ref{sec:kra17}, the {\it Gaia} parallax of DR~Tau
differs significantly from the parallaxes of other known members.
We retain  it in our catalog of members for the purposes of this
study, but its membership should be reassessed with the proper
motion and parallax from the next data release of {\it Gaia}.

In Table~\ref{tab:motion}, we list the  409
previously known members of Taurus
and the 18 new members presented in this study (Section~\ref{sec:spec}).
Binaries that are not resolved in any of our photometric catalogs
(Section~\ref{sec:phot}) appear as a single entry.

\section{Identification of Candidate Members of Taurus}
\label{sec:can}

\subsection{Proper Motions}
\label{sec:pm}

\subsubsection{Spitzer}
\label{sec:pm:irac}

A large portion of Taurus has been imaged with the Infrared Array Camera
\citep[IRAC;][]{faz04} on the {\it Spitzer Space Telescope}.
Those data have been previously used to classify the circumstellar disks
of the known members \citep{har05,luh06,luh10,gui07,esp14}
and to search for new disk-bearing members \citep{luh06,luh09a,luh09b,reb10}.
Because the IRAC images have been obtained at multiple epochs that span nearly
a decade, they can be used to search for new members based on proper motions.
These data are well-suited for detecting low-mass brown dwarfs in Taurus
given that such objects are brightest at IR wavelengths and extinction
is low in the IRAC bands.

IRAC contains four 256$\times$256 arrays and 
four broad-band filters centered at 3.6, 4.5, 5.8, and 8.0~$\mu$m, 
which are denoted as [3.6], [4.5], [5.8], and [8.0].
Each array has a plate scale of $1\farcs2$ pixel$^{-1}$, corresponding
to a field of view of $5\farcm2\times5\farcm2$.
Point sources within the images have a FWHM of $1\farcs6$--$1\farcs9$
for [3.6]--[8.0]. Following its launch in August 2003, {\it Spitzer} was
initially cooled with liquid helium. That cryogenic phase of {\it Spitzer}
ended in May 2009 when the helium was depleted. 
IRAC has continued to operate with the [3.6] and [4.5] bands.

\begin{figure}[h]
\centering
\includegraphics[trim = 0mm 0mm 0mm 0mm, clip=true, scale=0.8]{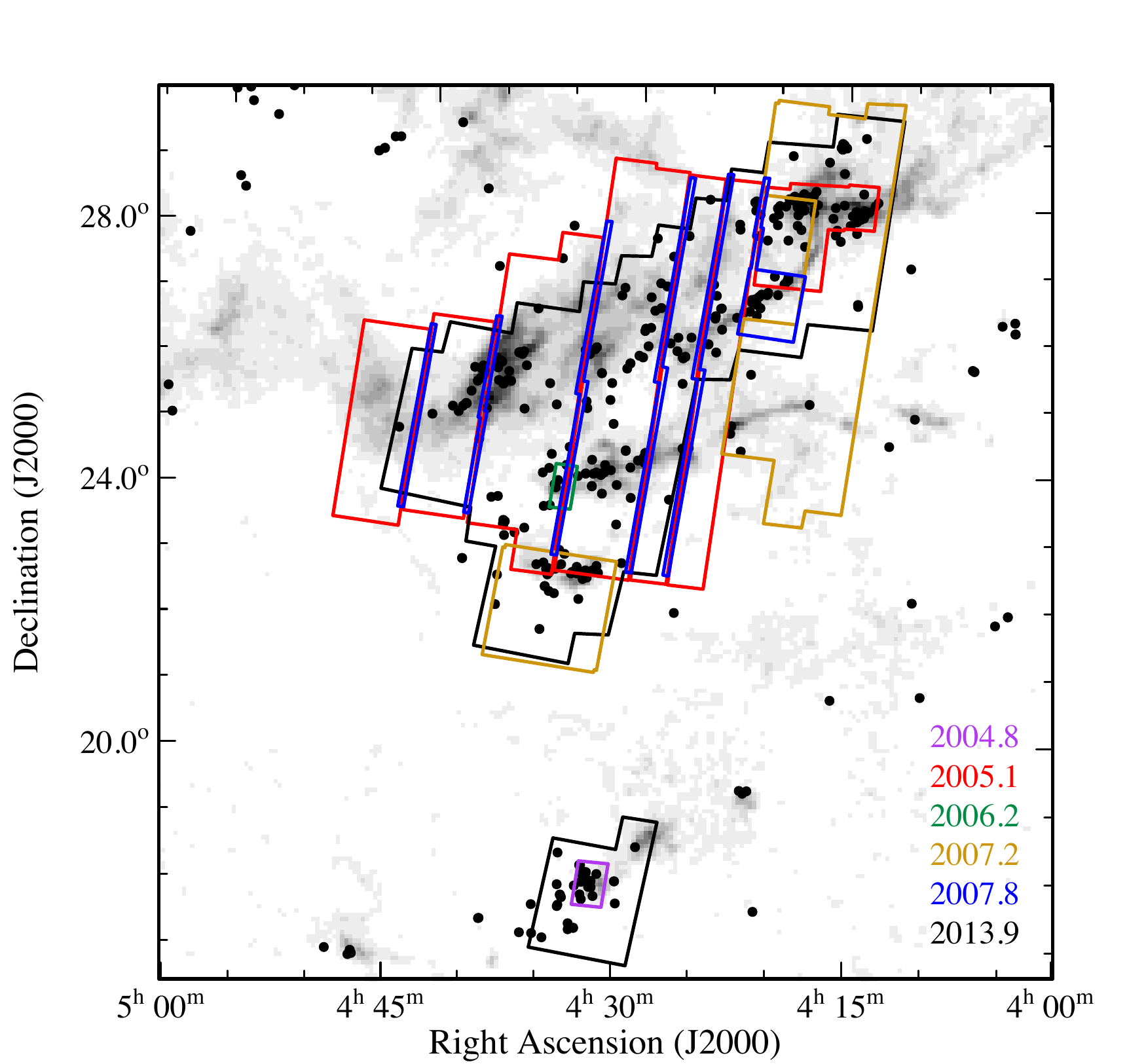}
\caption{
Map of the fields in the Taurus star-forming region that were imaged by IRAC
at multiple epochs (Table~\ref{tab:epochs}).
The known members of Taurus are indicated (filled circles).
The dark clouds in Taurus are displayed with a map of extinction
\citep[gray scale;][]{dob05}.
}
\label{fig:cov}
\end{figure}

We have retrieved from the {\it Spitzer} archive all [3.6] and [4.5] images
for areas that were imaged at multiple epochs that spanned several years.
In Table \ref{tab:epochs}, we list the Astronomical Observing Requests (AORs), 
program identifications (PIDs), and principle investigators (PIs) for these 
observations in Table \ref{tab:epochs}.  The spatial coverages of the six 
largest maps are shown in Figure \ref{fig:cov}.
Additional details regarding the observations from the cryogenic phase
(e.g., exposure times) are compiled by \citet{luh10}.
The one set of observations from the post-cryogenic phase
consists of mosaics in which the individual images have exposure times
of 10.4~s and dither steps of $100\arcsec$, resulting in nine frames per
position for a given band.

We measured astrometry for all sources in the IRAC images using
the methods described in \cite{esp16} and \cite{esp17}. In summary, we 
1) measured positions, fluxes ($F_\nu$), and signal-to-noise ratios (S/N's) 
using the point response function fitting routine 
in the Astronomical Point source EXtractor \citep[APEX;][]{mak05},
2) corrected those positions for distortion, 
3) iteratively refined the relative offsets and orientations between 
individual frames, and
4) measured astrometry for each source that was detected in at least three
frames among the [3.6] and [4.5] data.
Because the astrometry was unreliable for the brightest stars that were
near saturation, we rejected detections with
$F_\nu /({\rm exposure\ time})>0.73$ and $>$0.82~Jy/s in [3.6] and [4.5],
respectively.

For each IRAC source, a relative proper motion was
calculated by applying a linear fit to the measurements of right
ascension and declination as a function of time.
As done in \citet{esp17} for Chamaeleon~I,
we estimated the 25\%, 50\%, and 75\% quantiles in the errors in
$\mu_\alpha$ and $\mu_\delta$ as a function of S/N in the final epoch at
[3.6] by applying local
linear quantile regression with the function {\tt lprq} in the
R package {\it quantreg} \citep{koe16}.
At S/N$\geq$100, the median errors for both $\mu_\alpha$ and 
$\mu_\delta$ are 2.5~mas~yr$^{-1}$, and 50\% of those errors are between
1.9--3.7~mas~yr$^{-1}$.
The errors increase with lower S/N, reaching a median value of 25~mas~yr$^{-1}$
at S/N=3. At that S/N, 50\% of the errors are between 17--38~mas~yr$^{-1}$.
To minimize the number of contaminants while maximizing the number of known
Taurus members in our proper motion catalog,
we only considered motions in which the errors in both $\mu_\alpha$ and  
$\mu_\delta$ are $\leq$10~mas~yr$^{-1}$.
We also omitted measurements with S/N$<$5.5 and 7.0 in [3.6] and [4.5],
respectively, so that at least 25\% of sources above these limits in
S/N have errors less than 10~mas~yr$^{-1}$. The faintest objects that
satisfy these criteria have $[3.6]\sim17$ and $[4.5]\sim16.6$,
which corresponds to $K\sim18$ for members of Taurus.
In Table~\ref{tab:motion}, we list the resulting proper motions that
are available for {\color{red} 165} members of Taurus. 
Those motions are plotted in Figure \ref{fig:pm}, where we include
contours that represent all other sources with IRAC motions.

To design proper motion criteria for selecting candidate members of Taurus,
we began by estimating the intrinsic spread in proper motions within the Taurus
population using the highly accurate proper motions that are available for
16 known members from the first data release of the {\it Gaia} mission
\citep{gaia16a,gaia16b}. As shown in Figure~\ref{fig:pi},
those motions do exhibit a spread that is significantly larger than the errors
in the measurements, and most of them are contained within a radius of
10~mas~yr$^{-1}$.
Therefore, for each source of proper motions employed in our survey (IRAC,
UKIDSS, etc.), we identify candidate members based on proper motions that
have 1~$\sigma$ errors that overlap with the range of motions defined
by a radius of 10~mas~yr$^{-1}$ from the median value of the motions of
the known members. That threshold is indicated in Figure~\ref{fig:pm}
for each catalog of proper motions.
The IRAC proper motion of one known member, XEST 08-047, fails 
our criterion for selecting new candidates, but only by a small margin.
Its motion from \citet{roe10} is consistent with membership,
so we retain it as a member.
We note that \cite{bes17} measured a motion for 2MASS J04373705+2331080
that was discrepant from the known members by 2~$\sigma$, but the motion
measured with IRAC is consistent with membership.

\subsubsection{UKIDSS}

\begin{figure}[h]
\centering
\includegraphics[trim = 0mm 0mm 0mm 0mm, clip=true, scale=0.8]{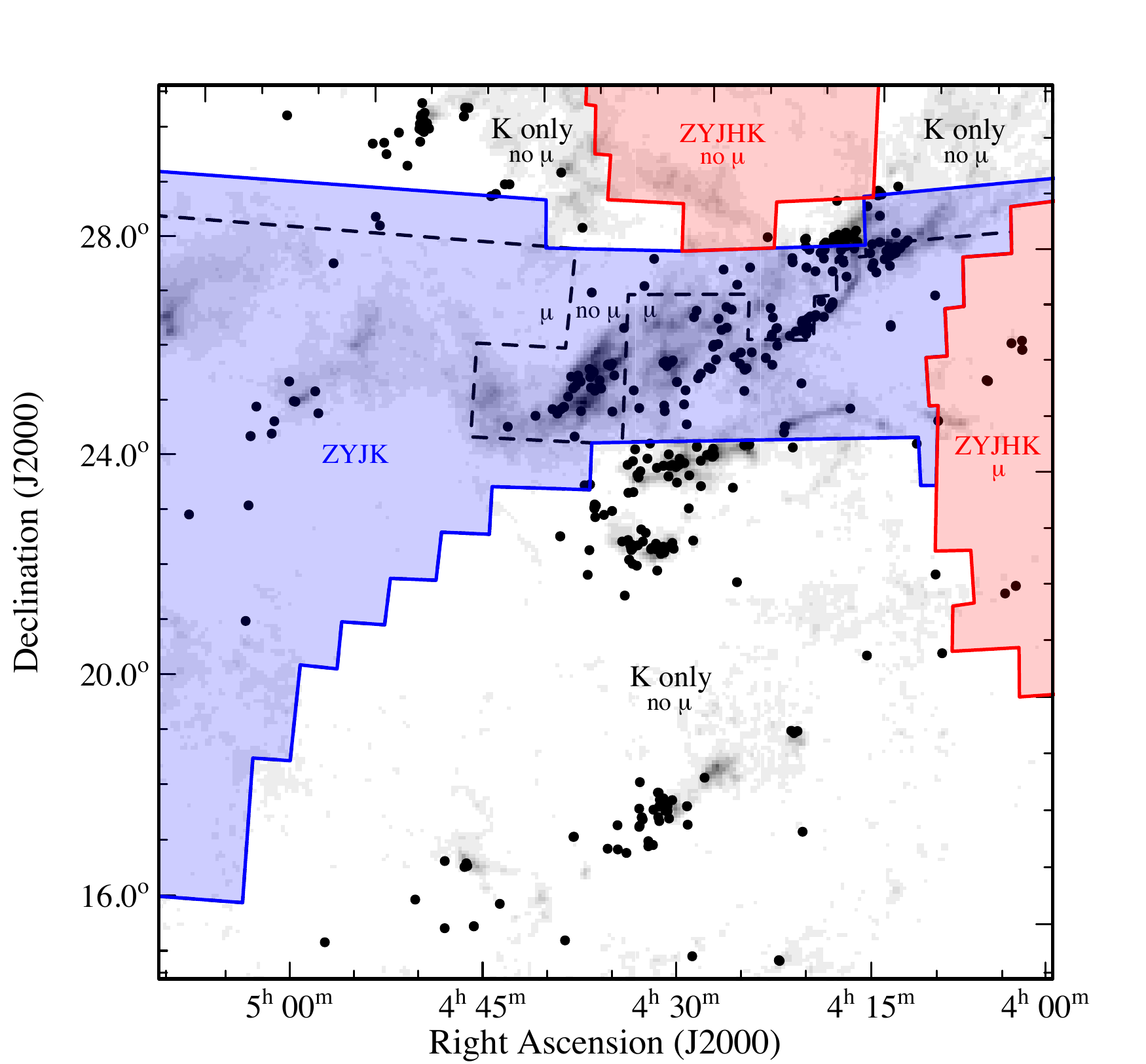}
\caption{
Map of the fields in the Taurus star-forming region that were imaged by UKIDSS.
Images in $K$ were obtained for the entirety of this map. The areas observed
in the other bands are outlined. We also mark the regions in which
UKIDSS proper motions are available. The known members of 
Taurus are indicated (filled circles). The dark clouds in Taurus are 
displayed with a map of extinction \citep[gray scale;][]{dob05}.
}
\label{fig:ukidss}
\end{figure}

The UKIDSS survey has imaged large sections of Taurus in five bands
($ZYJHK$), as illustrated in Figure~\ref{fig:ukidss}.
The $K$-band images exhibit S/N=10 at $K\sim17.5$, which is roughly
comparable to the IRAC images in terms of sensitivity to brown dwarfs
in Taurus. Some areas of Taurus were observed by UKIDSS at epochs that
spanned several years, which enabled the measurement of proper motions.
Therefore, we have made use of the proper motions from data release 10 of
UKIDSS that are available within a large area encompassing nearly all of the
known members of Taurus, which we defined as $\alpha=4^{\rm h}$ to
$5^{\rm h}10^{\rm m}$ and $\delta=15\arcdeg$ to $31\arcdeg$ (J2000).
As with the IRAC proper motions, we have excluded UKIDSS motions with errors
$>$10~mas~yr$^{-1}$ in $\mu_\alpha$ or $\mu_\delta$. We also have omitted
measurements for stars with $K_s<11.5$ because of saturation.
UKIDSS motions that satisfy those criteria are available for 17 known Taurus
members (includes new members).
Those motions are presented in Table~\ref{tab:motion} and
Figure~\ref{fig:pm}. They have median errors of $\sim$5~mas~yr$^{-1}$ in
$\mu_\alpha$ and $\mu_\delta$.  We examined the UKIDSS images of those
members for blending with neighboring stars, which might affect the proper 
motion measurements, but all of them appeared as unblended point sources.
One of these stars, 2MASS J04414825+2534304, has a UKIDSS motion that
is discrepant from the 10~mas~yr$^{-1}$ radius threshold in Figure~\ref{fig:pm}
by more than 1~$\sigma$, but its motions from IRAC and the combination of
2MASS and PS1 (Section~\ref{sec:pm:pan}) are consistent with membership.

\subsubsection{2MASS and {\it Gaia}}
\label{sec:2m.gaia}

\begin{figure}[h]
\centering
\includegraphics[trim = 0mm 10mm 0mm 0mm, clip=true, scale=0.9]{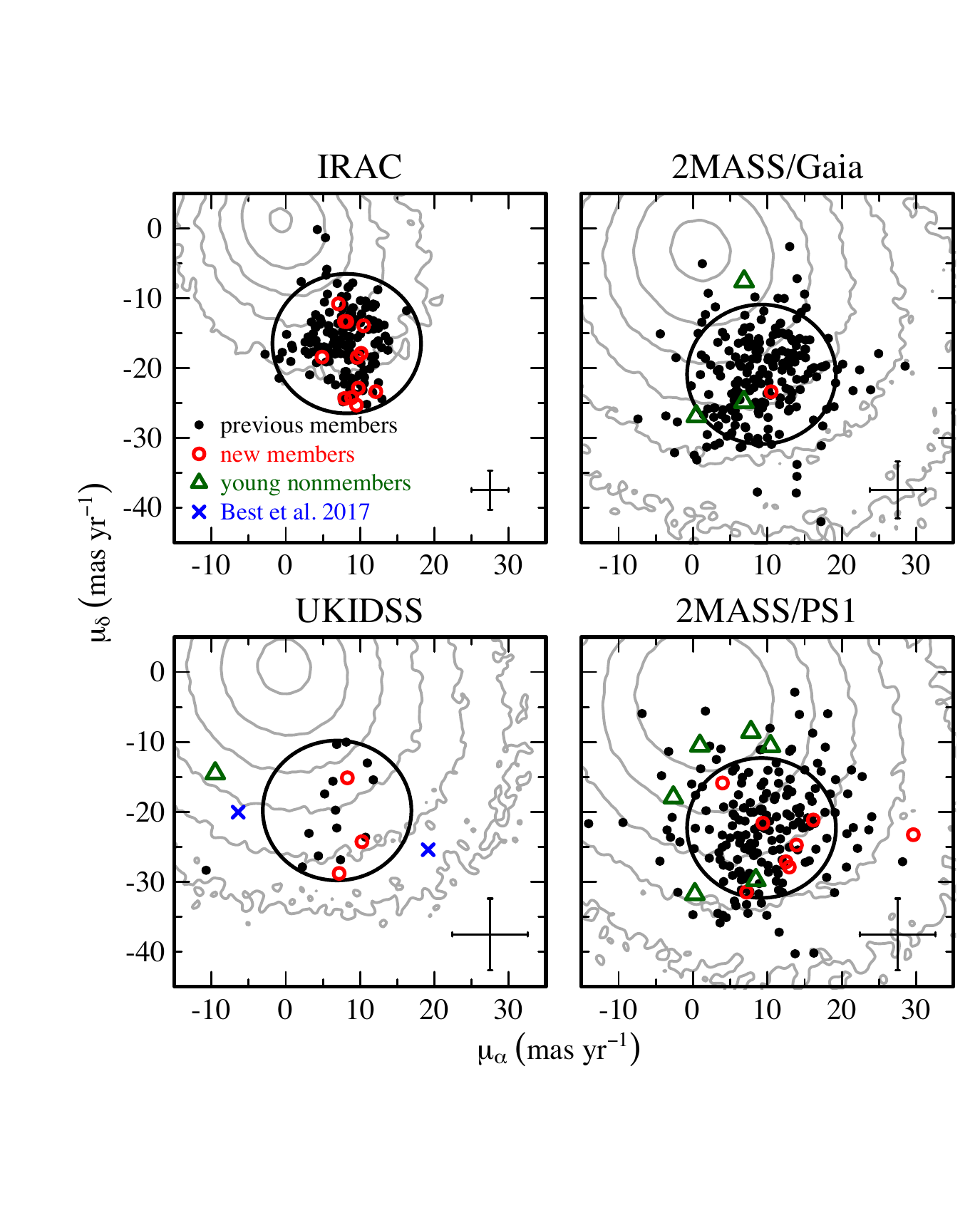}
\caption{
Relative proper motions measured with astrometry from IRAC, UKIDSS, 
2MASS/{\it Gaia}, and 2MASS/PS1 for the known members of Taurus (filled 
circles), new members from this work (open circles),
young objects found in our survey that do not appear to be members
(open triangles), and two candidate members from \cite{bes17} (crosses).
Measurements for other sources in the images of Taurus are represented by 
contours at log(number/(mas~yr$^{-1}$)$^2$)=1, 1.5, 2, 2.5, 3, and 3.5.
For each set of data, sources with 1~$\sigma$ errors that overlap with the
large circle are identified as proper motion candidates. The typical errors
are indicated in the corner of each diagram.
}
\label{fig:pm}
\end{figure}

In its first data release, the {\it Gaia} all-sky survey has
provided photometry in a broad optical band ($G$) and single-epoch positions
for stars with $G\lesssim20$ and measurements of parallaxes and proper 
motions for stars with $G\lesssim12$.
The former set of measurements reaches Taurus members as late as $\sim$M9
and as faint as $K\sim14$. 2MASS offers slightly better sensitivity to
members of Taurus and it was conducted more than a decade prior to {\it Gaia}.
Therefore, we can use proper motions based on the combination of
{\it Gaia} and 2MASS to help search for new substellar members of Taurus.
We have measured proper motions for sources
projected against Taurus (i.e., $\alpha=4^{\rm h}$--$5^{\rm h}10^{\rm m}$,
$\delta=15$--$31\arcdeg$) with the single-epoch positions from {\it Gaia}
and the astrometry from the 2MASS Point Source Catalog for sources detected
by both surveys. We have excluded measurements with errors $>$10~mas~yr$^{-1}$ 
in either component of the proper motion. In Figure~\ref{fig:pm}, we have 
indicated the 10~mas~yr$^{-1}$ radius threshold used for identifying 
candidate members based on the 2MASS/{\it Gaia} motions.

For each Taurus member that was detected by both {\it Gaia} and 2MASS,
we have inspected the 2MASS images to check whether the object was
blended noticeably with other stars. If it did not appear as a single
point source, we ignored its 2MASS/{\it Gaia} proper motion measurement.
We also discarded proper motions for members that were unresolved by 2MASS
but resolved as multiple objects by {\it Gaia}.  The resulting 
2MASS/{\it Gaia} motions for {\color{red} 210} known Taurus members
are included in Table~\ref{tab:motion} and Figure~\ref{fig:pm}.
For those members, the median errors in $\mu_\alpha$ and $\mu_\delta$
are $\sim$4~mas~yr$^{-1}$.
Thirteen members are discrepant by $>1$~$\sigma$ from the 10~mas~yr$^{-1}$
radius criterion in Figure~\ref{fig:pm}, but all of them have motions
from other catalogs that are consistent with membership.

\subsubsection{2MASS and Pan-STARRS1}
\label{sec:pm:pan}

The PS1 3$\pi$ survey obtained images of nearly all areas of sky at
$\delta\gtrsim-30\arcdeg$ in five bands \citep[$grizy_{P1}$,][]{ton12}
at several epochs between 2009 and 2014 \citep{cha16}.
The first data release of PS1 has provided photometry and astrometry measured
from coadditions of the multiple epochs in the 3$\pi$ survey.
The stacked images in the bands at the longest wavelengths are capable
of detecting members of Taurus at $K\lesssim15.5$, corresponding to spectral
types of early L. We have measured proper motions for sources projected
against Taurus using the astrometry from PS1 and 2MASS. To avoid saturated
detections, we only considered PS1 sources for which all bands were fainter
than 15 mag. As with the other sources of proper motions in our survey,
we have excluded 2MASS/PS1 measurements that have errors $>$10~mas~yr$^{-1}$.

We list 2MASS/PS1 motions for 195 known members of Taurus in 
Table~\ref{tab:motion}. We do not report measurements for members that
are blended noticeably with other stars in the 2MASS or PS1 images.
We retain measurements for members that are known to be multiple systems
but that are unresolved by 2MASS or PS1.
The values of $\mu_\alpha$ and $\mu_\delta$ in Table~\ref{tab:motion} have
median errors of slightly more than $\sim$4~mas~yr$^{-1}$.
The 2MASS/PS1 motions for known members are shown in Figure~\ref{fig:pm} with
the 10~mas~yr$^{-1}$ radius threshold for identifying new candidate members.
That criterion is not satisfied by 16 known members, but most of them
have motions in other catalogs that are consistent with membership.
The remaining source, 2MASS J04574903+3015195, is
discrepant by only slightly more than 1~$\sigma$ and shows strong
evidence of youth in its spectrum, so we retain it as a member for the
purposes of this study.

We note that {\it WISE} also has measured astrometry across our entire
survey field, but we did not include those data in our proper
motion measurements since the {\it WISE} images have lower resolution
than the other sources of astrometry that we have considered.

\subsection{Color-magnitude Diagrams}\label{sec:phot}

To further refine our candidate members of Taurus selected by proper motions
and to search for additional candidates that may lack such data, we have
constructed CMDs from several bands of optical and IR photometry. 
As with the proper motion measurements, we have considered photometry
for objects at $\alpha=4^{\rm h}$--$5^{\rm h}10^{\rm m}$ and
$\delta=15\arcdeg$--31$\arcdeg$.
These data consist of $G$ from the first data release of {\it Gaia},
$JHK_s$ from the 2MASS Point Source Catalog, $ZYJHK$ from data release
10 of UKIDSS, $W1$ and $W2$\footnote{{\it WISE} obtained images in bands at
3.5, 4.5, 12, and 22~$\mu$m, which are denoted as $W1$, $W2$, $W3$, and $W4$,
respectively.} from the AllWISE Source Catalog,
$riz$ from data release 13 of the Sloan Digital Sky Survey 
\citep[SDSS;][]{alb16}, $rizy_{P1}$ from the first data release of PS1 
\citep{fle16}, and [3.6] and [4.5] from all IRAC observations of Taurus prior 
to 2010, which were processed during the study of \citet{luh10}.
Additional bands are available from some of these catalogs, but they
are not sensitive to low-mass members of Taurus.
Some of the surveys offer multiple options of photometry for a given band.
We adopted the $1\arcsec$ radius aperture magnitudes from UKIDSS,
the point-spread-function (PSF) magnitudes from SDSS, and the
PSF magnitudes from the stacked images in PS1.
To avoid erroneous photometry of saturated stars, we did not use any PS1
measurements brighter than 15~mag and we excluded UKIDSS data at
$J<11$, $H<11.5$, and $K_s<10.5$. The $Z$ and $Y$ data from UKIDSS were
also omitted for $J<11$. For stars observed at two epochs in $K$ by UKIDSS,
we adopted the average of those measurements.

We identified all matches among the sources from the various catalogs.
When merging the photometry from different sources, we treated the following
filters as the same: $JHK_s$(2MASS)/$JHK$(UKIDSS) and $ri$(SDSS)/$ri_{P1}$.
When both 2MASS and UKIDSS data were available for a star, we adopted the
2MASS measurement if its error was $\leq$0.06. Otherwise, we adopted 
the UKIDSS photometry. If a star was detected in both SDSS and PS1 for
$r$ or $i$, we adopted the PS1 measurement.
The bands $Z$(UKIDSS)/$z_{P1}$ and $Y$(UKIDSS)/$y_{P1}$ are sufficiently
different that they are used separately in our CMDs. 
We found that the $z$ photometry from SDSS did not provide
significant added value in identifying candidates beyond the
data in $Z$(UKIDSS) and $z_{P1}$, so those measurements are not included
in our CMDs.

\begin{figure}[h]
\centering
\includegraphics[trim = 0mm 0mm 0mm 0mm, clip=true, scale=0.65]{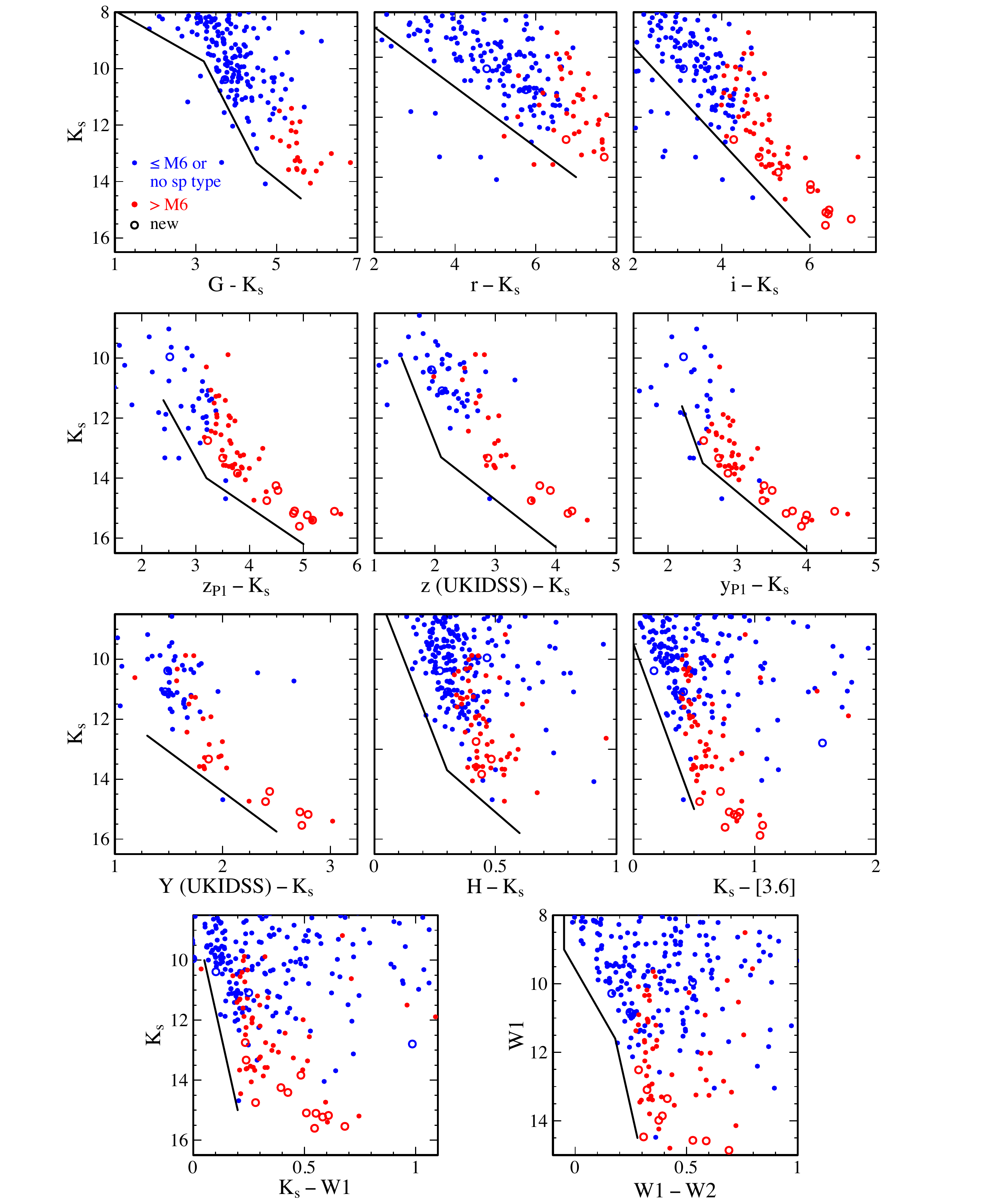}
\caption{
Extinction-corrected CMDs for previously known members of Taurus (filled 
circles) and new members from this work (open circles) based on photometry 
from {\it Gaia}, UKIDSS, PS1, 2MASS, {\it WISE}, and {\it Spitzer}. 
Among other stars detected in these surveys, we have selected candidate
members based on positions above the solid boundaries.
}
\label{fig:crit}
\end{figure}

In most of our CMDs, we have selected $K_s$ (or $K$) for the vertical axis
and have used colors that contain that band because it offers the greatest
sensitivity to low-mass members of Taurus (as well as low extinction)
among the available options. The one exception is a diagram of $W1$
versus $W1-W2$. As done in our previous surveys of this kind
\citep[e.g.,][]{luh03}, we have corrected the data in the CMDs for
extinction, which reduces contamination of background stars in
the areas of the diagrams inhabited by Taurus members.
We have estimated the extinction for each star by dereddening its
data in near-IR CMDs to the typical locus of young stars at the distance
of Taurus. In $H$ versus $J-H$, we define the locus as 
[$J-H=0.68$] for $H<13.0$ and 
[$J-H=0.128 \times H-0.984$] for $H\geq13.0$. Stars that lack
data in $H$ were dereddened in $K_s$ versus $J-K_s$ to a locus defined by 
 [$J-K_s=1$] for $K_s<14.5$ and 
[$J-K_s=0.3\times K_s-3.35$] for $K_s\geq14.5$.
For sources that lack the data necessary for either of those extinction
estimates, no correction is applied to their data.
For those stars, one could instead adopt the values inferred from extinction
maps \citep[e.g.,][]{dob05}, but such estimates would have large uncertainties
because of the low resolution of those maps, and they would be overestimated for
Taurus members, most of which are not subject to the total extinction through
the clouds.
Since the Taurus dark clouds cover only a small fraction of our survey field
(Figure~\ref{fig:new}), most stars in our CMDs exhibit negligible extinction
($A_J<0.1$). Near the clouds, the extinction estimates are concentrated
at $A_J<1$ and reach as high as $A_J\sim7$.
When correcting the photometry for extinction, we have adopted the reddening
relations from \citet{sch16}, $A_{3.6}=0.6$~$A_K$ 
\citep[][references therein]{xue16}, and $A_H/A_K=1.55$ \citep{ind05}.
The resulting extinction-corrected CMDs for Taurus are shown in
Figure~\ref{fig:crit}. In each CMD, we have selected a boundary that follows
the lower envelope of the sequence of known members. We have identified
stars as photometric candidate members if they appear above a boundary in
any diagram and do not fall below a boundary in any diagram. 

\section{Spectroscopy of Taurus Candidates}
\label{sec:spec}

\subsection{Observations}

We have pursued spectroscopy of an initial sample of our candidate members
of Taurus so that we can determine whether they are likely to be members
based on their spectral types and gravity-sensitive features.
During the selection of these targets, we gave higher priority to
candidates that were identified with both CMDs and proper motions,
appear in multiple CMDs, are located within a few degrees of known members,
and have CMD positions that are indicative of later spectral types.
For a given object, if the proper motion from one source supported
membership but the motion from another source was inconsistent with
membership, we still treated it as a viable candidate. 

We obtained near-IR spectra of 25 candidates with SpeX \citep{ray03}
at the NASA Infrared Telescope Facility (IRTF). The instrument was
operated in the prism mode with the $0\farcs8$ slit (0.7--2.5~$\mu$m, R=150).
We also observed nine additional candidates with the Gemini Near-Infrared
Spectrograph \citep[GNIRS;][]{eli06} in the cross-dispersed mode with the
31.7~l~mm$^{-1}$ grating and the $0\farcs67$ slit (0.8--2.5~$\mu$m, R=800).
The dates of these observations are listed in Table~\ref{tab:spec}.
The SpeX data were reduced with the Spextool package \citep{cus04} and
corrected for telluric absorption with the methods from \cite{vac03}.
The GNIRS data were reduced in a similar manner with routines in IRAF. 
To increase the S/N of the GNIRS spectra, we binned them in 25 pixel
increments, producing a resolution comparable to that of the SpeX data.

\subsection{Spectral Classifications}

\begin{figure}[h]
\centering
\includegraphics[trim = 0mm 10mm 0mm 0mm, clip=true, scale=1]{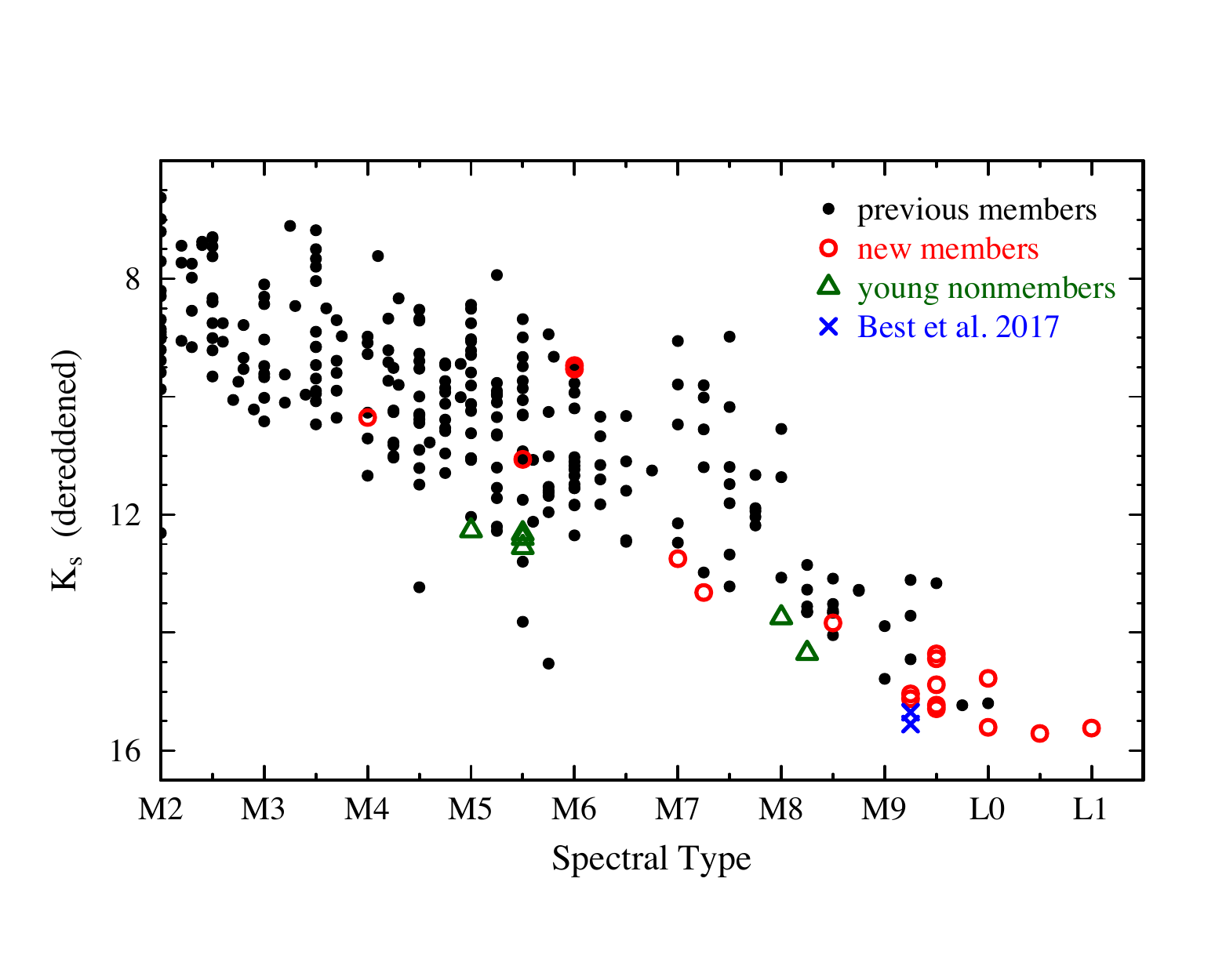}
\caption{
Extinction-corrected $K_s$ versus spectral type for the previously known members
of Taurus (filled circles), new members from this work (open circles),
young objects found in our survey of Taurus that do not appear to be members
(open triangles), and two candidate members from \citet{bes17} (crosses).
The members that appear below the sequence may be seen primarily in
scattered light because of an occulting circumstellar disk, which is
plausible given that they exhibit mid-IR excess emission.
}
\label{fig:kvspec}
\end{figure}

\begin{figure}[h]
\centering
\includegraphics[trim = 0mm 10mm 0mm 0mm, clip=true, scale=0.9]{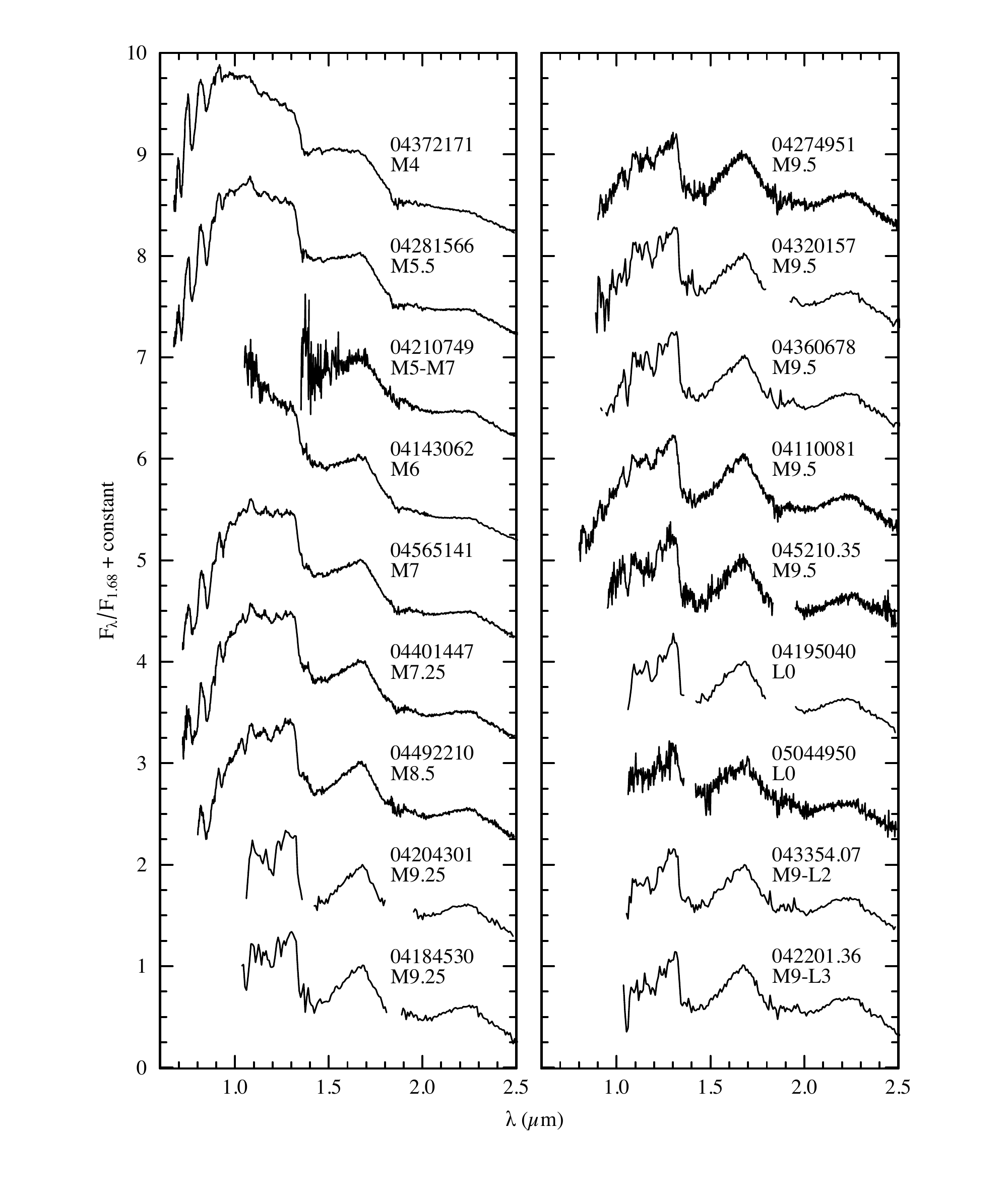}
\caption{
Near-IR spectra of new members of Taurus,
which have been dereddened to match standard young brown dwarfs \citep{luh17}.
These data have a resolution of $R=150$.
The data used to create this figure are available.
}
\label{fig:spec}
\end{figure}

We have used our near-IR spectra to measure spectral types for the
candidates and to assess whether they are likely to be members of Taurus.
Based on the photometry of these candidates, they should have
spectral types of mid-M or later if they are Taurus members.
Six candidates lack steam absorption bands, indicating that
they are earlier than $\sim$M0, and thus are likely reddened background stars.
For the remaining 28 candidates that do show those absorption bands, we
have assessed their ages, and thus their membership in Taurus, using 
spectral features that are sensitive to surface gravity, such as the shape
of the $H$-band continuum and the strength of the FeH band at 0.99~$\mu$m
\citep{luc01}. These objects were compared to both standards for field dwarfs 
\citep{cus05,ray09} and standards for ages of $\lesssim10$~yr \citep{luh17}. 
Four candidates are best matched by the field dwarf standards
and 24 candidates show evidence of youth. Eighteen of the young objects have
gravity-sensitive features that agree closely with those
of known members of Taurus \citep[][references therein]{luh17} and
have $K_s$ data that are consistent with other Taurus members near their
types (Figure~\ref{fig:kvspec}), so they are treated as confirmed members.
Their spectra are shown in Figure~\ref{fig:spec}. 

\begin{figure}[h]
\centering
\includegraphics[trim = 0mm 10mm 0mm 0mm, clip=true, scale=0.9]{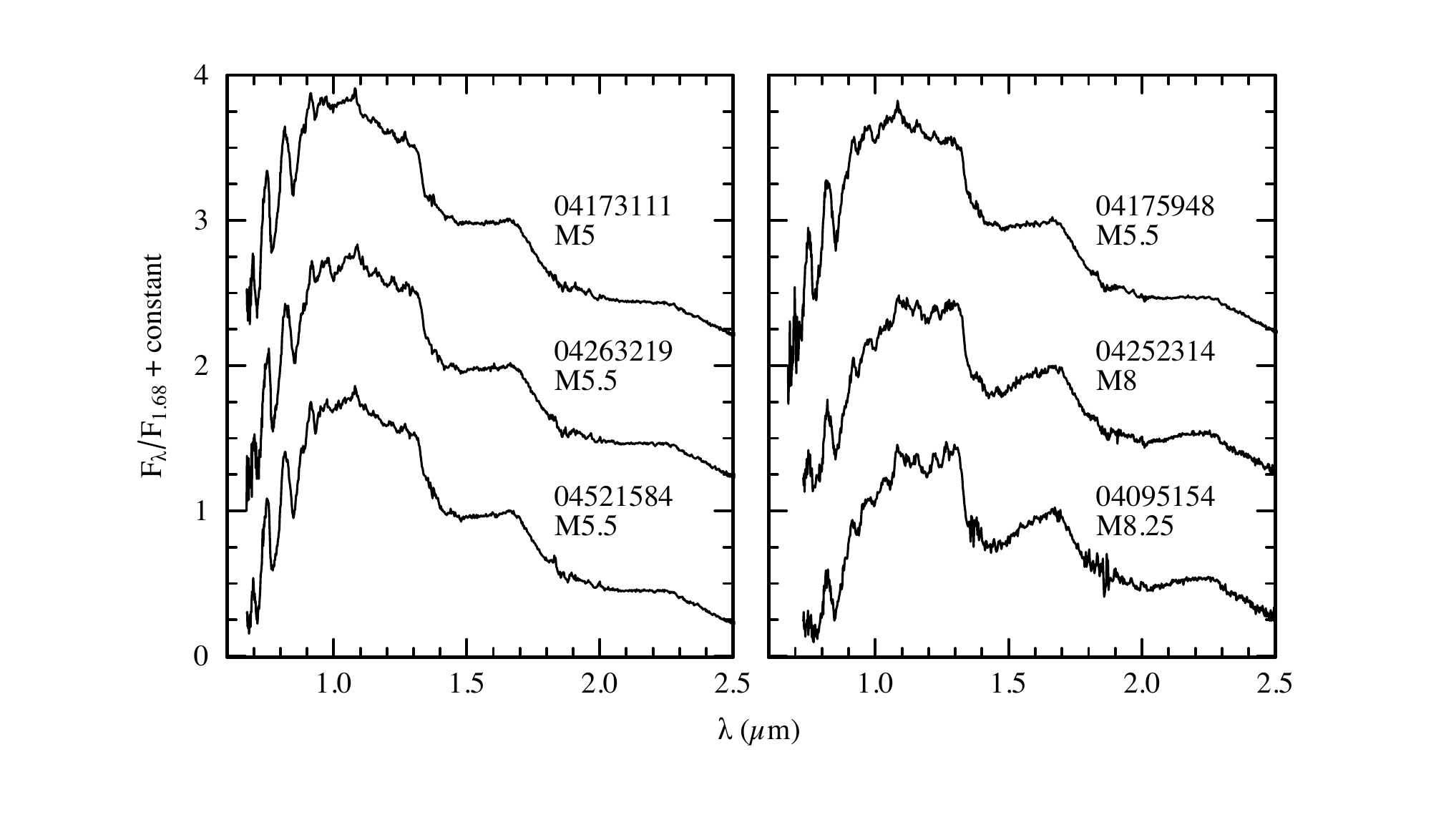}
\caption{
Near-IR spectra of objects from our survey that have spectral features
indicative of youth but that appear unlikely to be members of Taurus.
These data have been dereddened to match standard young brown dwarfs
\citep{luh17} and have a resolution of $R=150$.
The data used to create this figure are available.
}
\label{fig:yspec}
\end{figure}

We classify six of the young objects as likely non-members for the following
reasons. Four of the candidates exhibit less triangular $H$-band continua 
(2MASS J04252314+1735150, 2MASS J04095154+2000428) or stronger FeH absorption
(2MASS J04175948+2521283, 2MASS J04263219+1800280) than members of Taurus,
indicating somewhat older ages (see Figure~\ref{fig:yspec}).
All six candidates are unusually faint for Taurus members at their spectral
types, as shown in Figure~\ref{fig:kvspec}. None of them have mid-IR excess
emission that would indicate the presence of circumstellar material, so their
faint positions in that diagram cannot be explained by edge-on disks.
All but one of the six objects (2MASS J04095154+2000428) have proper motions
that are in the outskirts of the distribution of motions for known members
(Figure~\ref{fig:pm}), one of which fails the criteria used for selecting
candidates with UKIDSS motions.
Finally, all but one of those six candidates (2MASS J04175948+2521283) are
far from known members of Taurus (Figure~\ref{fig:new}).

Our classifications of the 34 spectroscopic targets are presented
in Table~\ref{tab:spec}.

Several of our new objects are among the faintest known members of Taurus,
as illustrated in the diagram of extinction-corrected $K_s$ versus spectral
type in Figure~\ref{fig:kvspec}. For instance, our sample includes the
four faintest known members and eight of the 10 faintest ones in
extinction-corrected $K_s$.
Assuming an age of 1~Myr and a $K_s$-band bolometric correction for
young late-M/early-L objects \citep{fil15}, the faintest new members
should have masses near 4--5~$M_{\rm Jup}$ according to evolutionary models 
\citep{bur97,cha00,bar15}.

\section{Circumstellar Disks Among New Taurus Members}
\label{sec:disks}

\cite{luh10} and \cite{esp14} presented the most recent compilations of
mid-IR photometry from {\it Spitzer} and {\it WISE} for the known members
of Taurus. They used those data to identify and classify circumstellar disks. 
 Roughly 2/3 of the known members exhibited mid-IR excess
emission that indicated the presence of a disk.
In their sample of members,
\cite{esp14} included the new members from \citet{luh17} with the
exception of 2MASS J04344586+2445145, which was mistakenly omitted.
In Table~\ref{tab:mid}, we have compiled {\it Spitzer}
and {\it WISE} data for that star, the three new members from \citet{abe14},
the three stars from \citet{kra17} that we have adopted as members, and the
new members from our study. One of our new members, UGCS J045210.35+303734.3,
is absent from Table~\ref{tab:mid} since it was not observed by {\it Spitzer}
and it was not detected by {\it WISE}.
For {\it Spitzer}, we have considered the four bands of IRAC and the
24~\micron\ band from the Multiband Imaging Photometer for Spitzer
\citep[MIPS;][]{rie04}, which is denoted as [24].
The {\it Spitzer} data were measured in the manner described by \cite{luh10}.
The {\it WISE} data are from the AllWISE Source Catalog.

\begin{figure}[h]
\centering
\includegraphics[trim = 0mm 0mm 0mm 0mm, clip=true, scale=0.9]{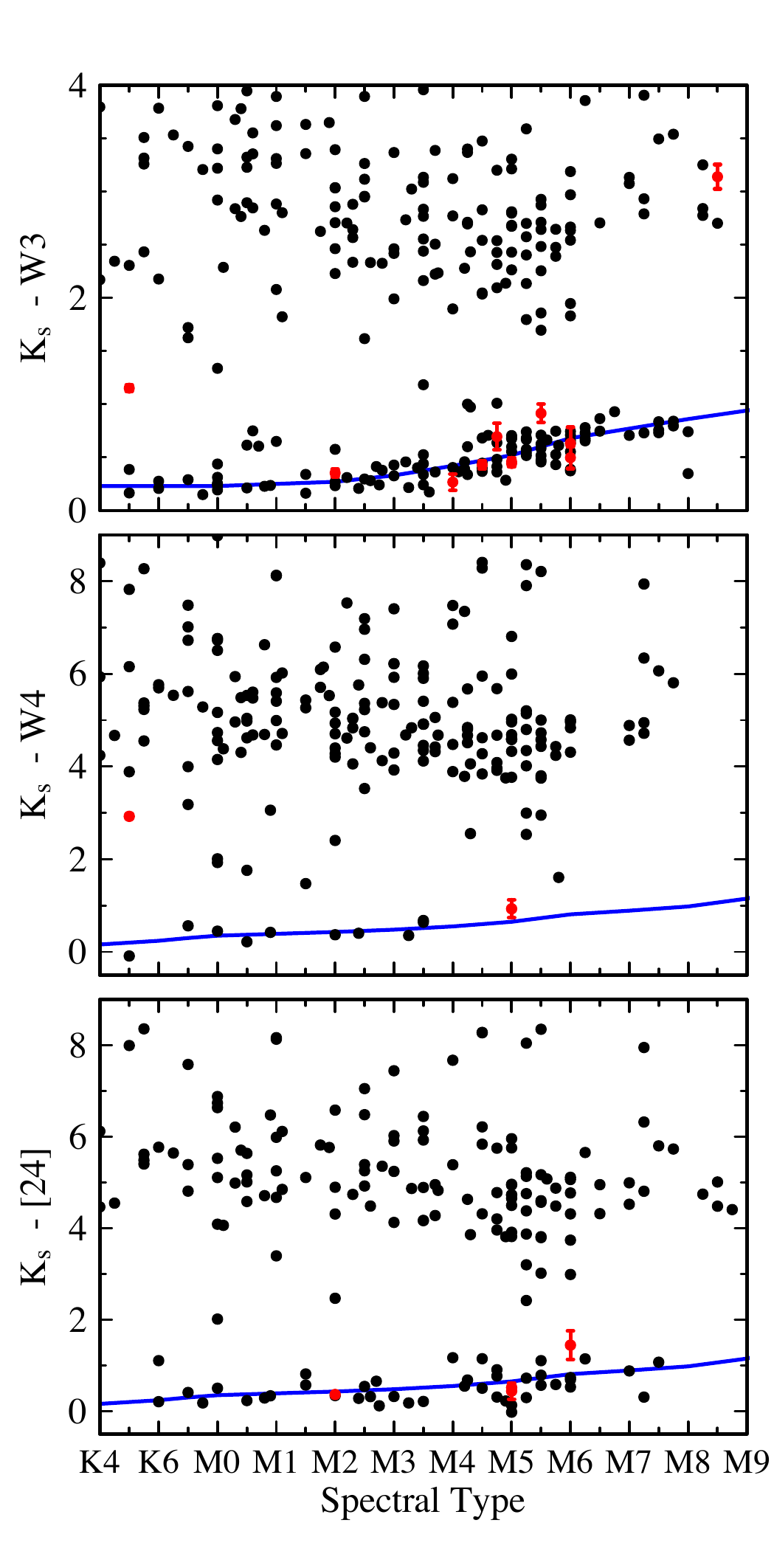}
\caption{
Extinction-corrected mid-IR colors as a function of spectral type 
for late-type members of Taurus (filled circles).
The members that have been adopted since \cite{esp14}
(Table~\ref{tab:mid}) are plotted in red and with the errors in their colors.
Two of these members exhibit significant
excess emission, which indicates the presence of circumstellar disks.
We mark the intrinsic photospheric colors for young objects (blue lines,
K. Luhman, in preparation). 
}
\label{fig:wisedisk}
\end{figure}

To determine whether the stars in Table~\ref{tab:mid} exhibit evidence of 
disks in their mid-IR photometry, we begin by plotting $K_s-W3$, $K_s-W4$, and
$K_s-[24]$ as a function of spectral type in Figure~\ref{fig:wisedisk} for the 
stars that have detections in $W3$, $W4$, or [24]. For comparison, we have
included previously known members of Taurus within that range of spectral
types and estimates for the typical intrinsic photospheric colors of young
stars (K. Luhman, in preparation). Among the sources in Table~\ref{tab:mid},
2MASS J04492210+2911124 (M8.5) and 2MASS J05080816+2427150 (K5) show
significant color excesses relative to photospheric colors in
Figure~\ref{fig:wisedisk}.
In their survey for disk-bearing members of Taurus using {\it WISE} data,
\citet{esp14} did not identify 2MASS J04492210+2911124 as a candidate because
its photometric error in $W3$ is slightly larger than the threshold that was 
adopted when selecting data to consider.
They did select 2MASS J05080816+2427150 as a disk candidate based on $W1-W4$,
but they found that it was heavily blended with another star in 
the {\it WISE} Atlas Image at $W4$, so that excess was ignored.
However, the Atlas Images were artificially blurred, whereas the
versions produced by \citet{lan14} and \citet{mei17} retain the intrinsic
resolution of the original data. In the $W4$ image of 2MASS J05080816+2427150
from the latter studies, the two stars are sufficiently resolved that we now
accept the $W4$ excess as reliable. As shown in Figure~\ref{fig:wisedisk},
2MASS J05080816+2427150 also has an excess in $W3$. Even the lower resolution
Atlas Image indicates that the $W3$ measurement should have little
contamination from the neighboring star, but that excess was not noticed by
\citet{esp14} because $W1-W3$ was just below their threshold for identifying
candidates.

\begin{figure}[h]
\centering
\includegraphics[trim = 0mm 0mm 0mm 0mm, clip=true, scale=0.85]{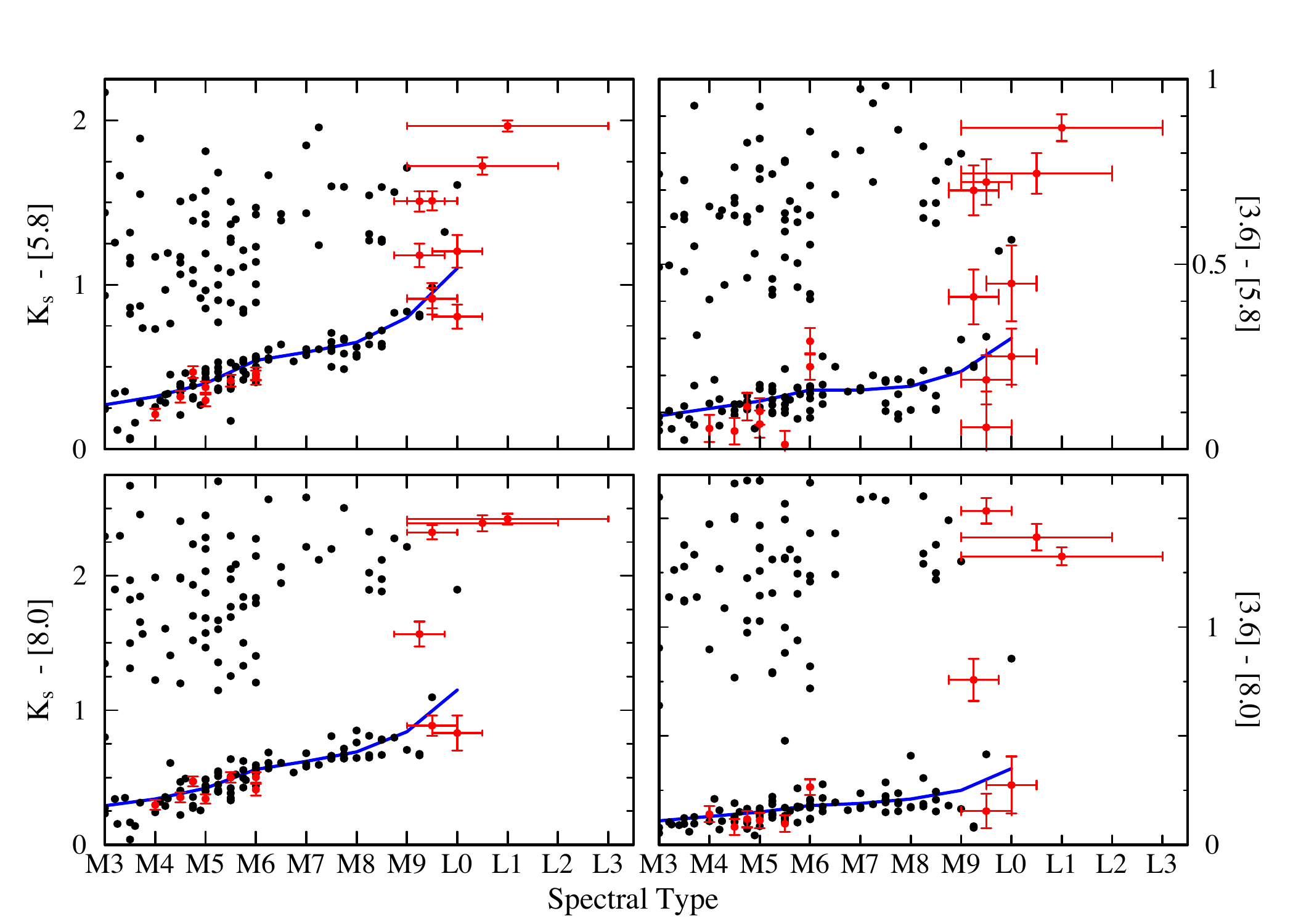}
\caption{
Extinction-corrected mid-IR colors versus spectral type for late-type members
of Taurus (filled circles). The members that have been adopted since 
\cite{esp14} (Table~\ref{tab:mid}) are plotted in red and with the errors in 
their colors. Uncertainties in spectral types are also included for the new
members that are later than M9.
We mark the intrinsic photospheric colors for young objects (blue lines,
K. Luhman, in preparation). 
Two of the new Taurus members with precise spectral types (at M9.25 and M9.5)
show clear excesses at [5.8] and [8.0], indicating that they have circumstellar
disks. Two additional new members are also redder than the photospheric
sequences at $\leq$L0, but since they have larger spectral type errors that
extend later than L0, and since the photospheric colors are ill-defined at
those types, we cannot determine conclusively whether color excesses from disks are present.
}
\label{fig:iracdisk}
\end{figure}

We have also examined the IRAC photometry in Table~\ref{tab:mid} for color
excesses. In Figure~\ref{fig:iracdisk}, we show $K_s-[5.8]$, $K_s-[8.0]$,
$[3.6]-[5.8]$, and $[3.6]-[8.0]$ as a function of spectral type for
objects from Table~\ref{tab:mid} that have detections in [5.8] or [8.0].
Among those sources with accurate spectral classifications,
2MASS J04204301+2810364 (M9.25) and 2MASS J04320157+1815229 (M9.5)
exhibit excesses at [5.8] and [8.0] relative to the photospheric colors.
Two of the members with larger errors in their types,
UGCS J043354.07+225119.1 (M9--L2) and UGCS J042201.36+265512.1 (M9--L3),
are significantly redder than stellar photospheres at $\leq$L0, but
since the photospheric colors are ill-defined at $>$L0 for ages of a few Myr,
it is unclear whether color excesses are present.
Additional data (i.e., detections at longer wavelengths)
are needed to determine if these two objects harbor circumstellar disks.

We comment briefly on the mid-IR colors of the
previously known L0 member 2MASS J04373705+2331080.
\citet{luh09b} found that it is only slightly redder than young L0 dwarfs
in the field, so they concluded that
its colors were probably entirely photospheric rather than containing
excesses from a disk. However, it does exhibit significant excesses
at both [5.8] and [8.0] relative to our adopted
photospheric sequences, which were estimated from the
bluest known members of Taurus, Chamaeleon~I, Upper Sco, IC~348, and
the TW Hya association. There are two possible explanations for those
red colors for 2MASS J04373705+2331080: it does have a disk, or
the photospheres of young $\gtrsim$L0 objects at a given age have a large spread
in their mid-IR colors. As with our new $\gtrsim$L0 members,
additional photometry at longer wavelengths is necessary for
determining whether 2MASS J04373705+2331080 has a disk.

\section{Spectroscopy of Perseus Candidates}\label{sec:pers}

In a recent survey for new members of IC~348 and NGC~1333 in Perseus,
\citet{luh16} used CMDs, proper motions, and other indicators of membership to
identify candidate members of the clusters. They obtained spectra of some of 
those candidates to measure their spectral types and determine if they were 
members. We have done the same for nine additional candidates from the CMDs in
\citet{luh16}. We have also performed spectroscopy on
source 48 from \citet{ots08}, which is a candidate companion to a
known member of NGC~1333, and [SVS76]~NGC~1333~7, which is a probable member of
NGC~1333 that was excluded from the census in \citet{luh16} because its
spectral classification was uncertain.

\begin{figure}[h]
\centering
\includegraphics[trim = 0mm 10mm 0mm 0mm, clip=true, scale=0.9]{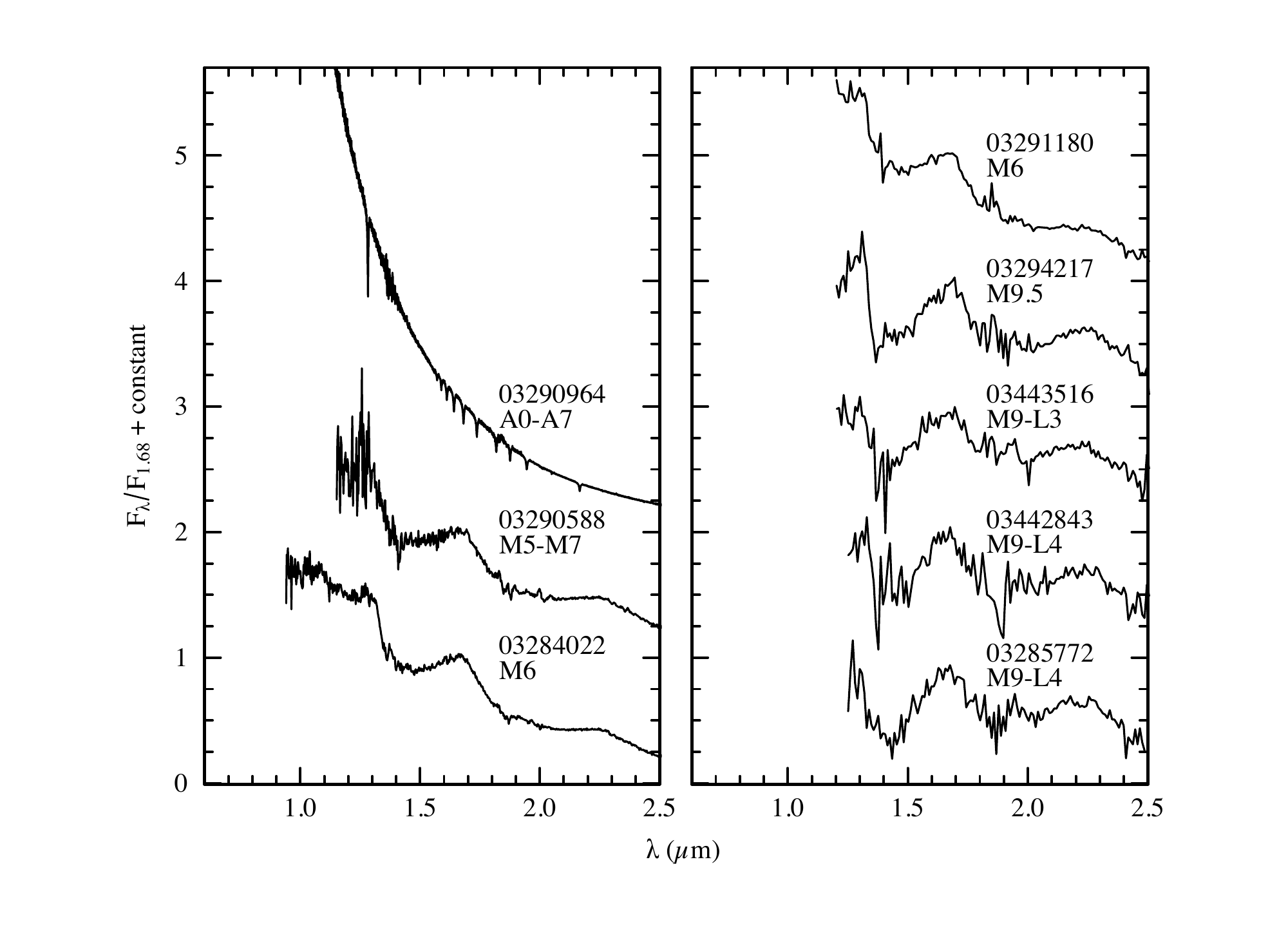}
\caption{
Near-IR spectra of new members of IC 348 and NGC 1333,
which have been dereddened to match standard young brown dwarfs \citep{luh17}.
These data have a resolution of $R=150$.
The data used to create this figure are available.
}
\label{fig:specpers}
\end{figure}

We obtained near-IR spectra of the 11 candidates in IC~348 and NGC~1333
with SpeX and GNIRS. The data were collected, reduced, and classified in
the same manner as the SpeX and GNIRS spectra of the candidates in Taurus
except for [SVS76] NGC 1333 7, which was observed with the SXD mode
of SpeX ($R=750$).
In Table~\ref{tab:perseus}, we list the spectral types, extinctions derived
from the spectra, membership assessments, spectrographs, and observing dates
for nine of the candidates. The remaining two candidates,
IC 348 IRS J03443631+3205066 and NGC 1333 IRS J03293317+3125495, are not
included in Table~\ref{tab:perseus} since the S/N's in their spectra
(from GNIRS) were too low for classification.
The spectra of LRL~595 and LRL~596 have insufficient S/N to definitively
determine whether they are young, but they are better matched by young objects
than field dwarfs, so we tentatively treat them as members of IC~348.
As indicated in Table~\ref{tab:perseus}, two and six candidates in
IC~348 and NGC~1333 are classified as members, respectively, which
results in a total of 480 and 209 known members when combined with the members
compiled by \citet{luh16}. We show the spectra of the eight new members in
Figure~\ref{fig:specpers}.

\begin{figure}[h]
\centering
\includegraphics[trim = 0mm 0mm 0mm 90mm, clip=true, scale=0.8]{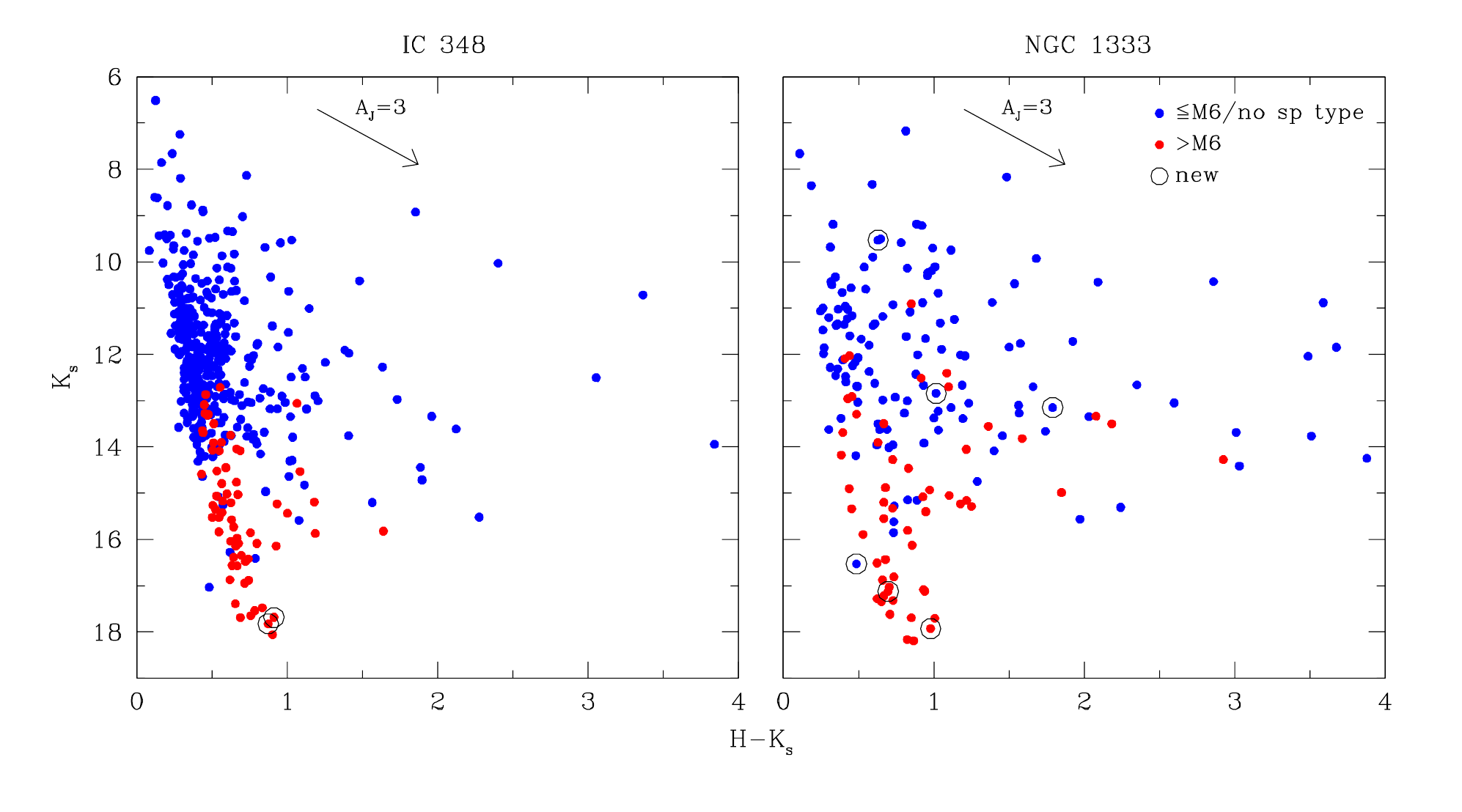}
\caption{
Near-IR CMDs for the previously known members of IC~348 and
NGC~1333 \citep[filled circles][]{luh16} and the candidates that we have
classified as new members (open and filled circles, Table~\ref{tab:perseus}).
}
\label{fig:cmdpers}
\end{figure}

To illustrate how the new members of IC~348 and NGC~1333 compare to
previously known members in terms of magnitude and reddening, we have
plotted diagrams of $K_s$ versus $H-K_s$ with the previously known and 
new members in Figure~\ref{fig:cmdpers}. Three of the new members of 
these clusters are among the faintest known members, and hence are
contenders for the least massive known members ($\sim5$~$M_{\rm Jup}$).
J03291180+3122036 is much fainter than other members of NGC~1333 near
its spectral type (M6), which indicates that it may be occulted by
a circumstellar disk and seen in scattered light. We cannot check
whether it exhibits the mid-IR excess emission expected from a disk
because it is projected against the bright extended emission that surrounds
[SVS76]~NGC~1333~3, which prevents a detection in images from {\it Spitzer}.

\section{Conclusions}

We have begun a survey for planetary-mass brown dwarfs in the Taurus
star-forming region, and we have continued a previous survey for
such objects in the IC~348 and NGC~1333 clusters in Perseus \citep{luh16}.
Our results are summarized as follows.

\begin{enumerate}

\item
\citet{luh17} recently presented a compilation of known members of Taurus.
We have applied revisions to that list by rejecting probable non-members
and adding stars that exhibit sufficient evidence
of membership from previous studies. While identifying stars to include
as members, we examined the 82 candidates from \citet{kra17} that were
absent from the list in \citet{luh17}. We have adopted three of those
candidates as members because they are similar to the known members
in terms of ages, kinematics, and distances (when available). 
Most of the candidates from \citet{kra17} that have {\it Gaia} parallaxes
and proper motions are distinct from the known members in those parameters,
indicating that they represent a different stellar population.
We have also considered two objects that were classified as L-type members of
Taurus by \citet{bes17}. We find that the spectra of both objects are
best matched by young standards at M9.25. Although they are clearly young,
neither of them shows strong evidence of membership, and they instead
may be intermediate-age ($>$10~Myr) non-members.

\item
We have identified candidate members of Taurus using proper motions and
photometry from a variety of sources ({\it Spitzer}, 2MASS, {\it Gaia}, PS1,
UKIDSS, SDSS, {\it WISE}). We have performed near-IR spectroscopy on
some of the more promising candidates, confirming 18 of them as new members.
They exhibit spectral types ranging from
M4 to early L, and they include the four faintest known members in
extinction-corrected $K_s$. The faintest new member should have a mass
of roughly 4--5~$M_{\rm Jup}$ according to evolutionary models.

\item
Two of the coolest new Taurus members (M9.25, M9.5) have mid-IR excess
emission, which indicates the presence of circumstellar disks.
Two additional members from our survey (M9--L2, M9--L3) also exhibit
red mid-IR colors relative to the photospheric values at $\leq$L0, but
it is unclear whether they have disks given the uncertainties in
their spectral types and the ill-defined nature of mid-IR colors of
young photospheres later than L0. 

\item
We have obtained near-IR spectra of candidate members of the IC~348
and NGC~1333 clusters in Perseus from \citet{luh16}.
Eight of these candidates are classified as new members, three of which
are among the faintest and least-massive known members ($\sim5$~$M_{\rm Jup}$).

\end{enumerate}

\acknowledgements
This work was supported by grant AST-1208239 from the NSF. We thank William
Best for providing his SpeX data for PSO~J060.3+25 and PSO~J077.1+24 and we
thank Lee Hartmann and Eric Mamajek for their comments on the paper.
The {\it Spitzer Space Telescope} is operated by JPL/Caltech under a contract
with NASA. The Gemini data were obtained through programs GN-2015B-FT-21,
GN-2015B-FT-27, GN-2016B-FT-8, and GN-2016B-FT-21.
Gemini Observatory is operated by AURA under a cooperative agreement with
the NSF on behalf of the Gemini partnership: the NSF (United States), the NRC
(Canada), CONICYT (Chile), the ARC (Australia),
Minist\'{e}rio da Ci\^{e}ncia, Tecnologia e Inova\c{c}\~{a}o (Brazil) and
Ministerio de Ciencia, Tecnolog\'{i}a e Innovaci\'{o}n Productiva (Argentina).
The IRTF is operated by the University of Hawaii under contract
NNH14CK55B with NASA.
The Pan-STARRS1 Surveys (PS1) and the PS1 public science archive have been made possible through contributions by the Institute for Astronomy, the University of Hawaii, the Pan-STARRS Project Office, the Max-Planck Society and its participating institutes, the Max Planck Institute for Astronomy, Heidelberg and the Max Planck Institute for Extraterrestrial Physics, Garching, The Johns Hopkins University, Durham University, the University of Edinburgh, the Queen's University Belfast, the Harvard-Smithsonian Center for Astrophysics, the Las Cumbres Observatory Global Telescope Network Incorporated, the National Central University of Taiwan, the Space Telescope Science Institute, the National Aeronautics and Space Administration under Grant No. NNX08AR22G issued through the Planetary Science Division of the NASA Science Mission Directorate, the National Science Foundation Grant No. AST-1238877, the University of Maryland, Eotvos Lorand University (ELTE), the Los Alamos National Laboratory, and the Gordon and Betty Moore Foundation.
The Center for Exoplanets and Habitable Worlds is supported by the
Pennsylvania State University, the Eberly College of Science, and the
Pennsylvania Space Grant Consortium.

\clearpage

\begin{deluxetable}{ll}
\tabletypesize{\scriptsize}
\tablewidth{0pt}
\tablecaption{Proper Motions of Taurus Members\label{tab:motion}}
\tablehead{
\colhead{Column Label} &
\colhead{Description}}
\startdata
Name & Source name\tablenotemark{a} \\
OtherNames & Other source names \\
IRACpmRA & IRAC relative proper motion in right ascension \\
e\_IRACpmRA & Error in IRAC\_pmRA \\
IRACpmDec & IRAC relative proper motion in declination \\
e\_IRACpmDec & Error in IRAC\_pmDec \\
UKIDSSpmRA & UKIDSS relative proper motion in right ascension \\
e\_UKIDSSpmRA & Error in UKIDSS\_pmRA \\
UKIDSSpmDec & UKIDSS relative proper motion in declination \\
e\_UKIDSSpmDec & Error in UKIDSS\_pmDec \\
2MGaiapmRA & 2MASS/{\it Gaia} proper motion in right ascension\\
e\_2MGaiapmRA & Error in 2MGaia\_pmRA \\
2MGaiapmDec & 2MASS/{\it Gaia} proper motion in declination\\
e\_2MGaiapmDec & Error in 2MGaia\_pmDec \\
2MPS1pmRA &2MASS/PS1 proper motion in right ascension \\
e\_2MPS1pmRA & Error in 2MPS1\_pmRA \\
2MPS1pmDec & 2MASS/PS1 proper motion in declination\\
e\_2MPS1pmDec & Error in 2MPS1\_pmDec 
\enddata
\tablecomments{This table is available in its entirety in a machine-readable form in the online journal.}
\tablenotetext{a}{Coordinate-based identifications from the 2MASS Point Source
Catalog when available. Otherwise, identifications are from the UKIDSS DR10 
Catalog or the AllWISE Source Catalog.}
\end{deluxetable}

\clearpage

\begin{deluxetable}{ccll}
\tabletypesize{\scriptsize}
\tablecaption{IRAC Observations of Taurus\label{tab:epochs}}
\tablehead{
\colhead{AOR} & \colhead{PID} & \colhead{PI} & \colhead{epoch}
}
\startdata
 3653120  & 6 & G. Fazio &   2004.8 \\
 3653376  & 6 & G. Fazio   & 2005.1 \\
 3653632  & 6 & G. Fazio  &  2005.1 \\
 3653888  & 6 & G. Fazio &   2005.1 \\
 3962880  & 37 & G. Fazio &   2005.1 
\enddata
\tablecomments{This table is available in its entirety in machine-readable form.}
\end{deluxetable}

\clearpage

\begin{deluxetable}{lllllll}
\tablecolumns{4}
\tabletypesize{\scriptsize}
\tablewidth{0pt}
\tablecaption{Spectral Types for Candidate Members of Taurus\label{tab:spec}}
\tablehead{
\colhead{Name} &
\colhead{Spectral Type\tablenotemark{a}} &
\colhead{A$_{J}$} &
\colhead{Instrument} & 
\colhead{Date}   &
\colhead{K$_{s}$} & 
\colhead{Ref}
}
\startdata
\cutinhead{New Members}
2MASS J04110081+2717163 & M9.5 & 0 & SpeX & 2017 Jan 14 & $14.36\pm0.06$ & 2MASS\\
2MASS J04143062+2807020 & M6 & 5.3 & SpeX & 2015 Dec 15 & $11.56\pm0.03$ & 2MASS\\
2MASS J04184530+2758484 & M9.25 & 0.15 & GNIRS & 2015 Nov 26 & $15.10\pm0.01$ & UKIDSS\\
2MASS J04195040+2820485 & L0 & 0 & GNIRS & 2015 Dec 31 & $15.61\pm0.02$ &  UKIDSS\\
2MASS J04204301+2810364 & M9.25 & 0.15 & GNIRS & 2015 Dec 5 & $15.18\pm0.01$ & UKIDSS\\
2MASS J04210749+2703022 & M5--M7 & 8.7 & SpeX & 2015 Dec 15 & $12.80\pm0.03$ & 2MASS\\
UGCS J042201.36+265512.1 & M9--L3 & 0--1.5 & GNIRS & 2016 Jan 1 & $15.84\pm0.02$ & UKIDSS\\
2MASS J04274951+2738155 & M9.5 & 0.29 & SpeX & 2015 Dec 15 & $14.55\pm0.01$ & UKIDSS\\
2MASS J04281566+2711110 & M5.5 & 0.06 & SpeX & 2017 Jan 14 & $11.09\pm0.02$ & 2MASS\\
2MASS J04320157+1815229 & M9.5 & 0.58 & GNIRS & 2016 Oct 1 & $15.11\pm0.01$ & UKIDSS\\
UGCS J043354.07+225119.1 & M9--L2 & 0--1.2 & GNIRS & 2016 Jan 3 & $15.88\pm0.02$ & UKIDSS\\
2MASS J04360678+2425500 & M9.5 & 0 & GNIRS & 2016 Oct 15 & $15.23\pm0.02$ & UKIDSS\\
2MASS J04372171+2651014 & M4 & 0.09 & SpeX & 2017 Jan 14 & $10.39\pm0.02$ & 2MASS\\
2MASS J04401447+2729112 & M7.25 & 0.03 & SpeX & 2017 Jan 14 & $13.33\pm0.04$ & 2MASS\\
2MASS J04492210+2911124 & M8.5 & 0 & SpeX & 2017 Jan 14 & $13.84\pm0.04$ & 2MASS\\
UGCS J045210.35+303734.3 & M9.5 & 0.29 & SpeX & 2017 Jan 14 & $15.40\pm0.01$ & UKIDSS\\
2MASS J04565141+2939310 & M7 & 0 & SpeX & 2017 Jan 14 & $12.75\pm0.02$ & 2MASS\\
2MASS J05044950+2510187 & L0 & 0.78 & SpeX  & 2017 Jan 13 & $15.07\pm0.01$ & UKIDSS\\
\cutinhead{Non-members}
2MASS J04095154+2000428 & young M8.25 & 0.09 & SpeX & 2017 Jan 14 & $14.38\pm0.05$ & 2MASS\\
2MASS J04143323+2806263 & $<$M0 & \nodata & SpeX & 2016 Jan 4 & $14.23\pm0.04$ & 2MASS\\
2MASS J04173111+2957305 & young M5 & 0.29 & SpeX & 2017 Jan 14 & $12.38\pm0.03$ & 2MASS\\
2MASS J04175948+2521283 & young M5.5 & 0.55 & SpeX & 2017 Jan 14 & $12.61\pm0.02$ & 2MASS \\
2MASS J04181651+2820036 & $<$M0 & \nodata & SpeX & 2015 Dec 15 & $13.97\pm0.04$ & 2MASS\\
2MASS J04183668+2723378 & $<$M0 & \nodata & GNIRS & 2016 Oct 9 & $15.69\pm0.02$ & UKIDSS\\
2MASS J04252314+1735150 & young M8 & 0 & SpeX & 2017 Jan 13 & $13.75\pm0.04$ & 2MASS\\
2MASS J04263219+1800280 & young M5.5 & 0.20 & SpeX & 2017 Jan 14& $12.63\pm0.02$ & 2MASS\\
2MASS J04294326+2429000 & $<$M0 & \nodata & SpeX & 2015 Dec 15 &$13.30\pm0.03$ & 2MASS\\
2MASS J04323814+2258149 & M0--M4V & \nodata & SpeX & 2015 Dec 15 &$14.28\pm0.01$ & UKIDSS\\
2MASS J04331507+2912364 & M4V & 0.44 & SpeX & 2017 Jan 14 & $12.28\pm0.02$ & 2MASS\\
UGCS J043529.68+240913.2 & $<$M0 & \nodata & GNIRS & 2016 Oct 15 & $15.65\pm0.02$ & UKIDSS\\
2MASS J04414141+2602092 & $<$M0 & \nodata & SpeX & 2015 Dec 15 & $13.84\pm0.06$ & 2MASS \\
2MASS J04521584+1517517 & young M5.5 & 0.15 & SpeX & 2017 Jan 14 & $12.38\pm0.03$ & 2MASS\\
2MASS J04550794+3004519 & L3V & \nodata & SpeX & 2017 Jan 13 & $15.47\pm0.01$ & UKIDSS\\
2MASS J04571245+2713065 & M4V & \nodata & SpeX & 2017 Jan 14 & $12.47\pm0.03$ & 2MASS
\enddata
\tablenotetext{a}{Uncertainties are $\pm0.5$~subclass unless indicated otherwise.}
\end{deluxetable}

\clearpage

\global\pdfpageattr\expandafter{\the\pdfpageattr/Rotate 90}
\clearpage

\begin{turnpage}
\begin{deluxetable}{lccccccccccc}
\tablecolumns{11}
\tabletypesize{\scriptsize}
\tablewidth{0pt}
\tablecaption{Mid-IR Photometry for Members of Taurus Adopted
Since \cite{esp14}\label{tab:mid}}
\tablehead{
\colhead{Name} &
\colhead{$W1$} &
\colhead{$W2$} &
\colhead{$W3$} &
\colhead{$W4$} &
\colhead{[3.6]} &
\colhead{[4.5]} &
\colhead{[5.8]} &
\colhead{[8.0]} &\colhead{[24]} & \colhead{Excess?} \\ 
&  \colhead{(mag)} &  \colhead{(mag)}  & \colhead{(mag)} & \colhead{(mag)} & \colhead{(mag)} & \colhead{(mag)} & \colhead{(mag)} & \colhead{(mag)} & \colhead{(mag)} 
}
\startdata
2MASS J04110081+2717163 & $13.92\pm0.03$ & $13.52\pm0.03$ & $12.37\pm0.52$\tablenotemark{a} & $>$8.94 & out & out & out & out & out & N\\
2MASS J04143062+2807020 & $10.92\pm0.02$ & $10.23\pm0.02$ & $10.15\pm0.10$ & $>$8.63 & $10.61\pm0.02$ & $10.24\pm0.02$ & $10.01\pm0.03$ & $10.04\pm0.03$ & out & N\\
2MASS J04184530+2758484 & $14.59\pm0.03$ & $14.00\pm0.06$ & $>$12.14 & $>$8.67 & $14.30\pm0.03$ & $14.10\pm0.03$ & $13.88\pm0.07$ & \nodata & \nodata & N\\
2MASS J04195040+2820485 & $15.06\pm0.04$ & $14.63\pm0.08$ & $>$12.07 & $>$8.00 & $14.85\pm0.03$ & $14.57\pm0.04$ & $14.40\pm0.10$ & \nodata & \nodata & N\\
2MASS J04204301+2810364 & $14.57\pm0.03$ & $14.04\pm0.06$ & $>$11.90 & $>$8.60 & $14.35\pm0.03$ & $13.94\pm0.03$ & $13.64\pm0.06$ & $13.58\pm0.09$ & \nodata & Y\\
2MASS J04210749+2703022 & $11.81\pm0.02$ & $10.79\pm0.02$ & $10.68\pm0.15$ & $>$8.51 & $11.24\pm0.02$ & $10.84\pm0.02$ & $10.52\pm0.03$ & $10.47\pm0.03$ & $9.04\pm0.31$ & N\\
UGCS J042201.36+265512.1 & $15.04\pm0.04$ & $14.32\pm0.08$ & $>$11.40 & $>$8.39 & $14.65\pm0.02$ & $14.23\pm0.02$ & $13.75\pm0.03$ & $13.30\pm0.04$ & \nodata & Y?\\
2MASS J04225416+2439538 & $8.55\pm0.03$ & $8.28\pm0.02$ & $8.15\pm0.02$ & $7.64\pm0.19$ & $8.37\pm0.02$ & $8.23\pm0.02$ & $8.28\pm0.03$ & $8.23\pm0.03$ & $8.04\pm0.06$ & N\\
2MASS J04244815+2643161  & $6.73\pm0.03$ & $7.30\pm0.02$ & $7.22\pm0.03$ & $7.18\pm0.20$\tablenotemark{a} & $7.45\pm0.02$ & $7.39\pm0.02$ & $7.38\pm0.03$ & $7.35\pm0.03$ & $7.24\pm0.04$ & N \\
2MASS J04274951+2738155 & $14.07\pm0.03$ & $13.68\pm0.04$ & $>$11.66 & $>$8.64 & $13.78\pm0.02$ & $13.60\pm0.02$ & $13.57\pm0.06$ & $13.61\pm0.08$ & \nodata & N\\
2MASS J04281566+2711110 & $10.84\pm0.02$ & $10.59\pm0.02$ & $10.16\pm0.08$ & $>$8.49 & $10.67\pm0.02$ & $10.61\pm0.02$ & $10.66\pm0.03$ & $10.57\pm0.03$ & \nodata & N\\
2MASS J04320157+1815229 & $14.55\pm0.03$ & $13.93\pm0.06$ & $11.40\pm0.34$\tablenotemark{a} & $8.37\pm0.52$\tablenotemark{a} & $14.23\pm0.03$ & $13.88\pm0.03$ & $13.47\pm0.06$ & $12.66\pm0.05$ & \nodata & Y\\
UGCS J043354.07+225119.1 & \nodata & \nodata & \nodata & \nodata & $14.83\pm0.02$ & $14.38\pm0.02$ & $14.06\pm0.05$ & $13.39\pm0.06$ & \nodata & Y?\\
2MASS J04344586+2445145 & $11.10\pm0.02$ & $10.87\pm0.02$ & $10.60\pm0.12$ & $>$8.81 & $10.94\pm0.02$ & $10.86\pm0.02$ & $10.82\pm0.03$ & $10.82\pm0.03$ & \nodata & N\\
2MASS J04354778+2523436 & $9.22\pm0.02$ & $9.00\pm0.02$ & $8.88\pm0.03$ & $8.68\pm0.47$\tablenotemark{a} & $9.04\pm0.02$ & $9.06\pm0.02$ & $8.98\pm0.03$ & $8.95\pm0.03$ & \nodata & N\\
2MASS J04355683+2352049 & $8.58\pm0.02$ & $8.47\pm0.02$ & $8.34\pm0.03$ & $8.23\pm0.32$\tablenotemark{a} & $8.54\pm0.02$ & $8.50\pm0.02$ & $8.44\pm0.03$ & $8.41\pm0.03$ & $8.33\pm0.04$ & N\\
2MASS J04360678+2425500 & $14.65\pm0.03$ & $14.26\pm0.05$ & $>$11.56 & $>$8.89 & $14.38\pm0.03$ & $14.13\pm0.03$ & $14.32\pm0.09$ & \nodata & \nodata & N\\
2MASS J04372171+2651014 & $10.28\pm0.02$ & $10.12\pm0.02$ & $10.10\pm0.08$ & $>$7.99 & $10.22\pm0.02$ & $10.10\pm0.02$ & $10.16\pm0.03$ & $10.07\pm0.03$ & \nodata & N\\
2MASS J04390453+2333199 & $9.48\pm0.02$ & $9.26\pm0.02$ & $9.15\pm0.03$ & $8.56\pm0.40$\tablenotemark{a} & $9.33\pm0.02$ & $9.24\pm0.09$ & $9.23\pm0.03$ & \nodata & $9.16\pm0.18$ & N\\
2MASS J04401447+2729112 & $13.09\pm0.03$ & $12.77\pm0.03$ & $11.76\pm0.30$ & $>$8.42 & out & out & out & out & out & N\\
2MASS J04492210+2911124 & $13.35\pm0.02$ & $12.94\pm0.03$ & $10.70\pm0.11$ & $8.57\pm0.43$ & out & out & out & out & out & Y\\
2MASS J04565141+2939310 & $12.51\pm0.02$ & $12.23\pm0.02$ & $12.01\pm0.34$ & $>$8.53 & out & out & out & out & out & N\\
2MASS J05044950+2510187 & $14.66\pm0.03$ & $14.32\pm0.05$ & $>$12.53 & $>$8.58 & $14.40\pm0.02$ & $14.30\pm0.03$ & $14.10\pm0.07$ & $14.08\pm0.13$ & out & N \\
2MASS J05080816+2427150 & $8.52\pm0.02$ & $8.29\pm0.02$ & $7.56\pm0.02$ & $5.77\pm0.05$ & out & out & out & out & out & Y
\enddata
\tablecomments{ 
Ellipses and ``out" indicate measurements that are absent because of non-detection
and a position outside of the camera's the field of view, respectively. 
}
\tablenotetext{a}{Detection is false or unreliable based on visual inspection.}
\end{deluxetable}
\end{turnpage}

\clearpage

\global\pdfpageattr\expandafter{\the\pdfpageattr/Rotate 360}

\begin{deluxetable}{lllllll}
\tabletypesize{\scriptsize}
\tablewidth{0pt}
\tablecaption{Spectral Types for Candidate Members of IC~348 and NGC~1333\label{tab:perseus}}
\tablehead{
\colhead{Name} &
\colhead{Other Names} &
\colhead{Spectral Type\tablenotemark{a}} &
\colhead{$A_J$} &
\colhead{Member?} &
\colhead{Instrument} &
\colhead{Date}}
\startdata
\cutinhead{IC~348}
IC 348 IRS J03442843+3211105 & NTC 08-202, LRL 595 & M9--L4 & 0--1.4 & Y? & GNIRS & 2016 Oct 9 \\
IC 348 IRS J03443516+3211052 & LRL 596 & M9--L3 & 0--1.3 & Y? & GNIRS & 2016 Oct 14 \\
IC 348 IRS J03444379+3213512 & LRL 22383 & mid-M? & 0.8 & N? &  GNIRS & 2016 Sep 28 \\
\cutinhead{NGC~1333}
2MASS J03284022+3125490 & \nodata & M6 & 3.0 & Y & SpeX & 2017 Jan 13 \\
NGC 1333 IRS J03285772+3118172 & [OTS2008] 19 & M9--L4 & 0--1.7 & Y & GNIRS & 2017 Jan 8 \\
NGC 1333 IRS J03290588+3116382 & [OTS2008] 48 & M5--M7 & 5.4 & Y & SpeX & 2017 Jan 13 \\
2MASS J03290964+3122564 & [SVS76] NGC 1333 7 & A0--A7 & 2.7 & Y & SpeX & 2017 Jan 13 \\
NGC 1333 IRS J03291180+3122036 & \nodata & M6 & 1.0 & Y & GNIRS & 2017 Jan 8 \\
NGC 1333 IRS J03294217+3117205 & \nodata & M9.5 & 0.1 & Y & GNIRS & 2017 Jan 8 
\enddata
\tablenotetext{a}{Uncertainties are $\pm0.5$~subclass unless indicated otherwise.}
\end{deluxetable}

\clearpage

\end{document}